\documentclass[structabstract]{aa}  
\usepackage{graphicx}
\usepackage{epstopdf}
\usepackage{epsfig}
\usepackage{tablefootnote}
\usepackage{color}
\usepackage{multirow}
\usepackage{rotating}
\usepackage{lscape}
\usepackage{longtable}
\usepackage{lipsum}
\usepackage{txfonts}
\usepackage{amssymb}
\usepackage{pdflscape}
\usepackage{lipsum, babel}
\usepackage{multicol}
\usepackage{subfig}

\begin{document}
   \title{New insights into time series analysis}
   \subtitle{II - Non-correlated Observations}

   \author{C. E. Ferreira Lopes$^{1,2,3}$ \and N. J. G.~Cross$^1$}
   \institute{
          SUPA (Scottish Universities Physics Alliance) Wide-Field Astronomy Unit, Institute for Astronomy, School of Physics and Astronomy, University of Edinburgh, Royal Observatory, Blackford Hill, Edinburgh EH9 3HJ, UK\\
          \and Departamento de F\'isica, Universidade Federal do Rio Grande do Norte, Natal, RN, 59072-970 Brazil \\
          \and National Institute For Space Research (INPE/MCTI), Av. dos Astronautas, 1758 – São José dos Campos – SP, 12227-010, Brazil  \\  \email{ferreiralopes1011@gmail.com}        }

   \date{Received November, 2016; accepted xxx, 2016}

 
  \abstract
   {Statistical parameters are used to draw conclusions in a vast number of fields such as
   finance, weather, industrial, and science. These parameters are
   also used to identify variability patterns on photometric data to 
   select non-stochastic variations that are indicative of astrophysical effects. New, 
   more efficient, selection methods are mandatory to analyze the huge amount of
   astronomical data.}
   {We seek to improve the current methods used to select
   non-stochastic variations on non-correlated data.}
   {We used standard and new data-mining parameters to analyze non-correlated data
   to find the best way to discriminate between stochastic and non-stochastic
   variations. A new approach that includes a modified Strateva function was performed to
   select non-stochastic variations. Monte Carlo simulations and public 
   time-domain data were used to estimate its accuracy and performance.
   }
   {We introduce 16 modified statistical parameters covering different features 
   of statistical distribution such as average, dispersion, and shape parameters. 
   Many dispersion and shape parameters are unbound parameters, i.e. equations 
   that do not require the calculation of average. Unbound parameters are
   computed with single loop and hence decreasing running time. Moreover, the 
   majority of these parameters have lower errors than previous parameters, which is mainly observed 
   for distributions with few measurements. A set of non-correlated
   variability indices, sample size corrections, and a new noise model 
   along with tests of different apertures and cut-offs on the data (BAS approach) are
   introduced.
   The number of mis-selections are reduced by about $520\%$
   using a single waveband and $1200\%$ combining all wavebands.
   On the other hand, the even-mean also improves the correlated indices
   introduced in Paper 1 \cite{FerreiraLopes-2016I}. The mis-selection rate is
   reduced by about $18\%$ if the even-mean is used instead of the mean to compute the correlated 
   indices in the WFCAM database.
   Even-statistics allows us to  improve the effectiveness of both correlated
   and non-correlated indices. }
   {The selection of non-stochastic variations is improved
   by non-correlated indices. The even-averages provide a better estimation of mean and 
   median for almost all statistical distributions analyzed. The
   correlated variability indices, which are proposed in the first paper of this series, are also improved if the even-mean is used.
   The even-parameters will also be useful for classifying light curves in the
   last step of this project.  We consider that the first step of this project,
   where we set new techniques and methods that provide a huge improvement on the
   efficiency of selection of variable stars, is now complete. Many of these 
   techniques may be useful for a large number of fields. Next, we will commence a new step of this project regarding
   the analysis of period search methods.}


   \maketitle
\section{Introduction}	

Statistical analysis is a vital concept in our lives because it is used to
understand what's going on and thereby enable us to make a decision. These kind of analyses are also used to assess 
theoretical models by experiments that are limited by experimental factors, 
leading to uncertainty. Measurements are usually performed many times to
increase the confidence level. The results are summarized by statistical 
parameters in order to communicate the largest amount of information as simply 
as possible. Statisticians commonly describe the observations by averages (e.g. 
arithmetic mean, median, mode, and interquartile mean), dispersion 
(e.g. standard deviation, variance, range, interquartile range, and absolute
deviation), shape of the distribution (e.g. skewness and kurtosis), and a
measure of statistical dependence (e.g. Spearman's rank correlation
coefficient). These parameters are used in finance, weather, industry,
experiments, science and in several other areas to characterize probability 
distributions. New insights on this topic should be valuable in 
natural sciences, technology, economy, and quantitative social science research.

Improvements on data analysis methods are mandatory to analyze the huge amount
of data collected in recent years. Large volumes of data with potential
scientific results are left unexplored or delayed owing to current inventory tools
that are unable to produce clear samples. In fact, we risk wasting the potential of a large 
part of these data despite efforts that have been undertaken \citealt[e.g.][]{vonNeumann-1941,vonNeumann-1942,Welch-1993,Stetson-1996,Enoch-2003,Kim-2014,Sokolovsky-2017}. Current
techniques of data processing can be improved considerably. For instance, the 
flux independent index that we proposed in a previous paper reduces the
mis-selection of variable sources by about $250\%$
\citep[][]{FerreiraLopes-2016I}. A reliable selection of astronomical databases
allows us put forward faster scientific results such as those enclosed in many
current surveys \citep[e.g.][]{Kaiser-2002,Udalski-2003,Pollacco-2006,
Baglin-2007,Hoffman-2009,Borucki-2010,Bailer-Jones-2013,Minniti-2010}. The 
reduction of misclassification at the selection step is crucial to follow up 
the development on the instruments themselves.

\begin{figure*}[htb]
 \centering
 \includegraphics[width=0.95\textwidth,height=0.8\textwidth]{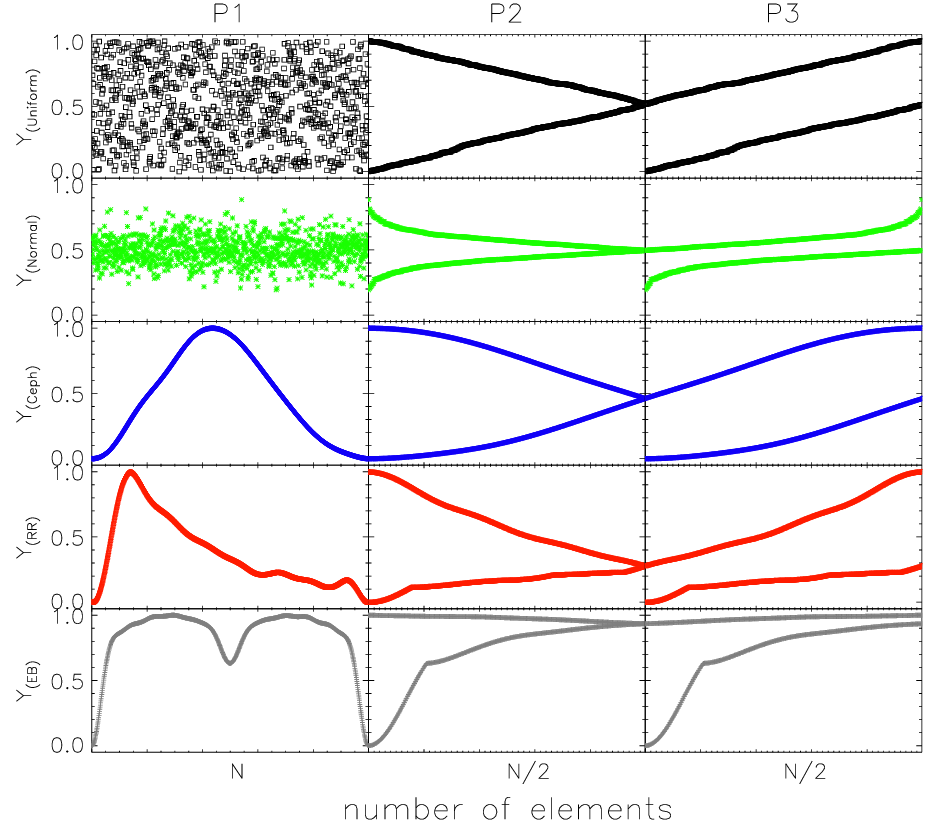}
  \caption{Uniform (black squares), normal (green asterisk), Ceph (blue
  diamond), RR (red triangle) and EB (grey plus) distributions with $1000$ 
  measurements as a function of the number of elements. The same distributions 
  are showing with different arrangements (see Sect.~\ref{sec_Notation} for more details).}
 \label{fg_dist}
\end{figure*}

The current project discriminates between correlated and
non-correlated observations to set the best efficiency for selecting
variable objects in each data set. Moreover,
\citet[][]{FerreiraLopes-2016I} establishes criteria that allow us to compute
confidence variability indices if the interval between measurements used to
compute statistical correlations is a small fraction of the variability
periods and the interval between correlated groups of observations.
On the other hand the confidence level of statistical parameters
increases with the number of measurements. Improvements to statistical
parameters for which there are few measurements are crucial to analyze surveys
 such ase PanSTARRS, with a mean of about $12$ measurements
in each filter, and the extended VVV project (VVVX), between $25$ to $40$ measurements
\citep[][]{Chambers-2016,Minniti-2010}.
\citealt[][]{Sokolovsky-2017} tested $18$ parameters ($8$ scatter-based
parameters  and $10$ correlated-based parameters), comparing their performance.
According to the authors the correlation-based indices are more efficient in
selecting variable objects than the scatter-based indices for data sets
containing hundreds of measurement epochs or more. The authors proposed a
combination of interquartile range \citep[IQR - ][]{Kim-2014} and the von
Neumann ratio  \citep[$1/\eta$ -][]{vonNeumann-1941,vonNeumann-1942} as a 
suitable way to select variable stars. A maximum interval of two days for
measurements is used to set those used to compute the correlated
indices. This value is greater than the limit required to compute 
well-correlated measurements. Moreover its efficiency should take into account the
number of well-correlated measurements instead of simply the number of epochs.
Indeed, maybe the authors did not account for our correlated indices 
\citep[][]{FerreiraLopes-2016I} because that would require a bin of shorter
interval than the smallest variability period. This allows us to get accurate
correlated indices for time series with few correlated measurements. The 
constraints and data used by \citealt[][]{Sokolovsky-2017}
limit a straightforward comparison between correlated and non-correlated 
indices performed by the authors. Therefore, the approach used to perform the 
variability analysis may only be chosen after examining the time interval among
the measurements (see Sect. \ref{sec_thebestway}).

Statistical parameters such as standard deviation and kurtosis as function of 
magnitude have been used as the primary way to select variable stars 
\citep[e.g.][]{Cross-2009}. This method assumes that for the same magnitude
stochastic and non-stochastic variation have different statistical properties. 
To compute all current dispersion and almost all shape parameters the averages, 
must also be calculated and thus we increase the uncertainties and 
processing time. Indeed, statistical properties still exist even where averages
are unknown. In this fashion, \citet[][]{Brys-2004} proposed a robust measure of
skewness called the 'medcouple' through a comparison of quartiles and pairs of
measurements that allow us to compute these parameters without use averages. However, 
 the medcouple measure has a long running time since the number of possible combination increases by
factorial of the number of measurements. However, we can use a similar idea to
propose new averages, dispersions, and shape parameters that have a smaller
running time.

This work is the second in a series about new insights into time series
analysis. In the first paper we assessed the discrimination of variable stars from
noise for correlated data using variability indices
\citep[][]{FerreiraLopes-2016I}. In this work, we analyze new statistical parameters and their
accuracy in comparison with previous parameters to increase the
capability to discriminate between stochastic and non-stochastic distributions. We also
look into their dependence with the number of epochs to determine statistical
weights to improve the selection criteria. Lastly we use a noise model to 
propose a new non-correlated variability index. Forthcoming papers will
study how to use the full current inventory of period finding methods to clean 
the sample selected by variability indices.

The notation used is described in Sect.~\ref{sec_Notation} and next we suggest
a new set of statistics in Sect.~\ref{sec_Even_Statistic}. In
Sections~\ref{sec_Sumary_tests} and \ref{sec_noise_model} the new parameters
are tested and a new approach to model the noise and and select variable stars
is proposed. Next, the selection criteria are tested on real data in
Sect.~\ref{sec_real_data}. Finally, we summarize and provide our conclusions in
Sect.~\ref{sec_Conclusions}.

\section{Notation}\label{sec_Notation}

Let $Y{'} := y{'}_{1} \leq y{'}_{2} \leq \ldots \leq y{'}_{c} \leq \ldots \leq
y{'}_{N'}$  from where the kernel function is defined by

\begin{equation}
  Y:=\left\{\begin{matrix}
  y_{i} \in Y' & \forall & y_{i} = y{'}_{i} & \, {\rm if} \quad N^{'} \quad {\rm even} \\ 
  y_{i} \in Y' & \forall & y_{i} \neq y{'}_{\rm Int(N'/2)+1}  & \, {\rm if} \quad N^{'} \quad {\rm odd} 
  \end{matrix}\right.
\label{eq_defvec}
\end{equation}

\noindent where $\rm Int(N'/2)$ means the integer part (floor) of half the number of
measurements. The lower contribution to compute statistical parameters for
symmetric distributions is given by $y{'}_{\rm Int(N'/2)+1}$ since it is the nearest
measurement to the average. The value $Y$ is a sample that has an even number of
measurements (N) that also can be discriminated into the subsamples $Y^{-}$ and
$Y^{+}$ composed of measurements $y_{i} \leq y_{N/2}$ and $y_{i} > y_{N/2}$, respectively. Different arrangements can be taken into consideration, such as

\begin{itemize}
 \item[\textbf{$P1$}] - unsystematic $y_{i}$ values;
 \item[\textbf{$P2$}] - increasing order of $Y^{-}$ and $Y^{+}$; 
 \item[\textbf{$P3$}] - decreasing order of $Y^{-}$ and increasing order of
 $Y^{+}$;
\end{itemize}

\noindent where the measurement of $Y^{-}$ and $Y^{+}$ for $P2$ and $P3$ assume 
the same position on the x-axis only to provide a better display. 

Five distributions were used to test our approach. These distributions were generated
to model both variable stars and noise. Uniform and normal distributions that
can mimic noise were generated by the \textit{IDL} function \textit{RANDOMU}.
This function returns pseudo-random numbers thar are uniformly distributed and randomly
drawn from a multivariate normal distribution, where a mean value of $0.5$
and full width at half maximum (FWHM) equal to $0.1$ were assumed for the normal
distributions to provide a range of values from about $0$ to $1$. On
the other hand we generate Cepheid (Ceph), RR-Lyrae (RR), and
eclipsing-binary (EB) distributions that are similar to 
typical variable stars. Ceph, RR, and EB models were based on the OGLE light curves
\textit{OGLE LMC-SC14 109671}, \textit{OGLE LMC-SC21 59535}, and \textit{OGLE LMC-SC2 180186}, respectively.
These models were generated in two steps: first a harmonic fit
was used to create a model and, second, these distributions were
sampled at random points to get the measurements.

Figure~\ref{fg_dist} shows the $P1-3$ arrangements for uniform (black
squares), normal (green asterisks), Ceph (blue diamonds), RR (red triangles) and
EB (grey plus signs) distributions. The colours and symbols used in this diagram
were adopted throughout the paper to facilitate their identification.
The symmetry found in the $P2$ and $P3$ yields unique information about the shape
and  dispersion parameters. Therefore, the measurements of $Y^{-}$ and $Y^{+}$ are 
combined to propose a new set of statistical parameters (see Table~\ref{tb_varind_new}).

\section{Even-statistic (E)}\label{sec_Even_Statistic}	

Non-parametric statistics are not based on probability distributions whose interpretation does not depend on the fitting of
parametrized distributions. The typical parameters used for descriptive and 
inferential purposes are the mean, median, standard deviation, skewness, 
 and kurtosis, among others. These parameters are defined to be a function of a 
sample that has no dependency on a parameter, i.e. its values are the same for 
any type of arrangement, for instance $P1-3$ (see Fig.~\ref{fg_dist}). All 
dispersion and almost all shape parameters are dependent on some kind of
average, i.e. they describe distributions around average values. Indeed, 
the dispersion and shape parameters still exist even where the averages are unknown. 
For instance, the standard deviation and absolute deviation provide an 
estimate of dispersion about the mean value.
Therefore, we propose the even-statistic as an alternative tool to
assess dispersion and shape parameters based only on the measurements, i.e.
unbound estimates.

The most appropriate to compute shape parameters is with even number of measurements
if we consider that the same number of measurements should appear on
both sides of the distribution for appropriate comparison. For instance, 
consider a distribution with an even number of measurements, where the mean
value can be computed as $\sum^{\rm Int(N'/2)}_{i=0}y^{'}_{i}/N' + 
\sum^{N}_{\rm i=Int(N'/2) + 1}y^{'}_{i}/N'$ and  the first and second term 
are the weights of the left and right sides of the distributions with the same 
number of elements. On the other hand, for odd numbers of measurements, the 
weight for both sides of the distribution are not equivalent. The single 
measurement that can be withdrawn to correct such variation is $y{'}_{\rm
Int(N/2)+1}$  since if we withdraw any of the other $y{'}_{i}$ measurements we
would increase the difference. The kernel given by Eq.~\ref{eq_defvec} provides
samples with even numbers of measurements and hence the same weight for both
sides. The parameters proposed with Eq.~\ref{eq_defvec} are called `even 
parameters'. Moreover, the even number of measurements allows us to compare 
single pairs of measurements among $Y^{-}$ and $Y^{+}$ and therefore to estimate 
dispersion and shape values without the average into account. The new 
statistical parameters to compute averages (see
Sect.~\ref{sec_Central_Tendencies}),  dispersions (see 
Sect.~\ref{sec_Dispersion_Parameters}), and shapes (see Sect.~\ref{sec_Shape_Parameters}) are described below.

\begin{table*}[htbp]
 \caption[]{Variability statistical analyses in the present
 work.}\label{tb_varind_new}
 \centering    
 \begin{tabular}{l l l l }        
 \hline\hline                 
 N & Statistic & Definition & Reference \\    
 \hline     \\                    
 
   1 & Even-mean    & $EA_{\mu} = \frac{1}{N}\sum_{i=0}^{N} y_{i}$ & Average  \vspace{0.25cm}\\
   2 & Even-median & $EA_{m} = \frac{y_{\frac{N}{2}} + y_{\frac{N}{2}+1}}{2}$ & Average  \vspace{0.25cm}\\

   3 & Even-mean standard deviation &  $ED_{\sigma_\mu} = \sqrt{\frac{ 1 }{ \left( N'-1 \right) } \sum_{i=1}^{N'} \left( y'_{i} - EA_{\mu} \right)^{2}} $ & Dispersion  \vspace{0.25cm}\\
   4 & Even-median standard deviation &  $ED_{\sigma_m} = \sqrt{\frac{ 1 }{ \left( N'-1 \right) } \sum_{i=1}^{N'} \left( y'_{i} - EA_{M} \right)^{2}} $ & Dispersion  \vspace{0.25cm}\\     
   5 & Even-mean-absolute deviation &  $ED_{\mu} = \frac{ 1 }{ N' } \sum_{i=1}^{N'} \left| y'_{i} - EA_{\mu} \right| $ & Dispersion  \vspace{0.25cm}\\
   6 & Even-median-absolute deviation &  $ED_{m} = \frac{ 1 }{ N' } \sum_{i=1}^{N'} \left| y'_{i} - EA_{M} \right| $ & Dispersion  \vspace{0.25cm}\\       
   7 & Even-absolute deviation & $ED = \frac{ 2 }{ N }\sum_{i=1}^{N/2} \left( y_{N-i} - y_{i} \right)$ & Dispersion \vspace{0.25cm}\\
   8 & Even-deviation (1) & $ED_{(1)} = \sqrt{ \frac{1}{N-1}\sum_{i=1}^{N/2}  \left( y_{N-i} - y_{i}\right)^{2} }$ & Dispersion \vspace{0.25cm}\\
   9 & Even-deviation (2)  & $ED_{(2)} = \sqrt{ \frac{1}{N-1}\sum_{i=1}^{N/2}  \left(y_{\frac{N}{2}+i} - y_{i}\right)^{2} }$ & Dispersion  \vspace{0.25cm}\\
   10 & Even Interquartile range  & ${\rm  EIQR} = EA_{m}(Y^{+}) - EA_{m}(Y^{-})$ & Dispersion  \vspace{0.25cm}\\ 
    
   11 & Even-skewness & $ES =  \frac{ \frac{1}{N'} \sum_{i=1}^{N'} \left(y'_{i} - EA_{\mu}\right)^{3} }{  ED_{\sigma_\mu}^{3}  }$ & Shape  \vspace{0.25cm}\\
   12 & Even-kurtosis & $EK =  \frac{ \frac{1}{N'} \sum_{i=1}^{N'} \left(y'_{i} - EA_{\mu}\right)^{4} }{  ED_{\sigma_\mu}^{4} } - 3$  & Shape  \vspace{0.25cm}\\
   13 & Even-shape (1) & $ES_{(1)} =    \frac{ \frac{1}{N} \sum_{i=1}^{N/2} \left(y_{N-i} - y_{i}\right)^{3} }{  ED_{(1)}^{3}  } - 2.26$ & Shape  \vspace{0.25cm}\\
   14 & Even-shape (2) & $ES_{(2)} =   \frac{ \frac{1}{N} \sum_{i=1}^{N/2} \left(y_{\frac{N}{2}+i} - y_{i}\right)^{3} }{ ED_{(2)}^{3} } - 1.52$  & Shape  \vspace{0.25cm}\\
   15 & Even-shape (3) & $EK_{(1)} =   \frac{ \frac{1}{N} \sum_{i=1}^{N/2} \left(y_{N-i} - y_{i}\right)^{4} }{ ED_{(1)}^{4}   } - 6$ & Shape  \vspace{0.25cm}\\
   16 & Even-shape (4) & $EK_{(2)} =   \frac{ \frac{1}{N} \sum_{i=1}^{N/2} \left(y_{\frac{N}{2}+i} - y_{i}\right)^{4} }{  ED_{(2)}^{4}  } - 2.46$ & Shape  \vspace{0.25cm}\\

\hline                                   
 \end{tabular}
\end{table*}

\subsection{Averages (A)}\label{sec_Central_Tendencies}

Considering the kernel given by Eq.~\ref{eq_defvec}(see
Sect.~\ref{sec_Notation}), we propose new average statistics given by

\begin{equation}
  EA_{\mu} = \frac{1}{N}\sum_{i=0}^{N} y_{i},
  \label{eq_defmean} 
\end{equation}
 
\noindent and

\begin{equation}
  EA_{m} = \frac{y_{\frac{N}{2}} + y_{\frac{N}{2}+1}}{2}
  \label{eq_defmed} 
\end{equation}

\noindent where $EA_{\mu}$ and $EA_{m}$  are named as even-mean and even-median.
These expressions mimic the mean ($A_{\mu}$) and median ($A_{M}$). Moreover, 
$EA_{\mu} = A_{\mu}$ and $EA_{m} = A_{m}$ when the number of measurements ($N'$)
is an even number. A comparison between these measurements is performed in
Sect.~\ref{sec_Accurate}.

\subsection{Dispersion parameters (D)}\label{sec_Dispersion_Parameters}

Statistical dispersion is used to measure the amount of sample variance and it 
is computed using the absolute or square value of the distance between the 
measurements and the average. Improving the estimation of averages can provide 
better accuracy of dispersion parameters such as the mean standard deviation
$(D_{\sigma_\mu})$, median standard deviation $(D_{\sigma_m})$, mean absolute
standard deviation $(D_{\mu})$, and median absolute standard deviation
$(D_{M})$. Therefore we propose the even-dispersion parameters that are 
computed using the even-averages (see Table~\ref{tb_varind_new}; 
the even-mean standard deviation $(ED_{\sigma_\mu})$, the even-median standard deviation 
$(ED_{\sigma_m})$, the even-mean-absolute deviation $(ED_{\mu})$, and 
the even-median-absolute deviation $(ED_{m})$). The accuracy of these 
parameters is assessed in Sect.~\ref{sec_Accurate}. We caution that these
parameters return the same values as the previous parameters for even numbers of
measurements.

Moreover, using the single combination between measurements $Y^{-}$ with $Y^{+}$,
we also can estimate the amount of variation or dispersion of a sample. From the
kernel given by Eq.~\ref{eq_defvec} we propose the following  dispersion parameter, written
as

\begin{equation}
  ED = \frac{ 1 }{ N }\sum_{i=1}^{N/2} \left( y_{N-i} - y_{i} \right) = \frac{ 1 }{ N }\sum_{i=1}^{N/2} \left( y_{\frac{N}{2}+i} - y_{i} \right).
  \label{dabsdev} 
\end{equation}

The two parts are the same since $\sum_{i=1}^{N/2} y_{N-i} 
= \sum_{i=1}^{N/2} y_{\frac{N}{2}+i} $. The value $ED$ means even-absolute deviation
because such a sum is always positive, i.e.  $\left( y_{N-i} - y_{i}\right) 
\geq 0$ and $\left( y_{\frac{N}{2}+i} - y_{i}\right) \geq 0$.  Moreover,
a simple identity is found for distributions with an even number of
measurements, i.e.

\begin{eqnarray}
ED &=&  \frac{ 1 }{ N }\sum_{i=1}^{N/2} \left( y_{N-i} - y_{i}  + EA_{m} - EA_{m} \right)  \nonumber \\
        &=& \frac{1}{N}\sum_{i=1}^{N/2} \left( y_{N-i} - EA_{m}\right) - \left(y_{i}  - EA_{m}\right)  \nonumber  \\
        &=&  ED_{m} 
 \label{proofabsstd}        
\end{eqnarray}

\noindent since $\left( y_{N-i} - EA_{m}\right) \ge 0$ and $\left(y_{i} -
EA_{m}  \right) \le 0$. Indeed, we also can mimic the standard 
deviation by proposing two new even-dispersion parameters given by

\begin{equation}
  ED_{(1)} = \sqrt{\frac{1}{N-1}\sum_{i=1}^{N/2}  \left( y_{N-i} - y_{i}\right)^{2} }
  \label{eq_defd1} 
\end{equation}

\noindent and,  

\begin{equation}
  ED_{(2)} = \sqrt{\frac{1}{N-1}\sum_{i=1}^{N/2}  \left(y_{\frac{N}{2}+i} - y_{i}\right)^{2} }.
  \label{eq_defd2} 
\end{equation}

The even-dispersion parameters $ED$, $ED_{(1)}$, and $ED_{(2)}$ are
unbound, i.e. they are not dependent on the average. These parameters allow us describe the dispersion of a distribution instead of
the dispersion about an average. Moreover, a strict relationship between
$ED_{(1)}$ and $ED_{(2)}$ with $ED_{\sigma_m}$ is found when we have 
even numbers of measurements, i.e.

\begin{eqnarray}
 ED^{2}_{(1)} &=&  \frac{1}{N-1}\sum_{i=1}^{N/2} \left( y_{N-i} - y_{i} + EA_{m}
 - EA_{m} \right)^{2}  \nonumber \\ &=& \frac{1}{N-1}\sum_{i=1}^{N/2}\left[ \left(y_{N-i}  - EA_{m}  \right) - \left( y_{i} - EA_{m}\right) \right]^{2}  \nonumber  \\
        &=& \frac{1}{N-1}\sum_{i=1}^{N/2} \biggl[ \left(y_{N-i}  - EA_{m}  \right)^2 + \left( y_{i} - EA_{m}\right)^2  \nonumber \\
        & &  - 2\left(y_{N-i}  - EA_{m}  \right) \times \left( y_{i} - EA_{m}\right) \biggr] \nonumber \\
        &=& ED^{2}_{\sigma_m} - 2\times {\rm Cov}(y_{N-i},y_{i})  
 \label{eq_proofstd1}        
\end{eqnarray}

\noindent while for $ED_{(2)}$,

\begin{eqnarray}
 ED^{2}_{(2)} &=&  \frac{1}{N-1}\sum_{i=1}^{N/2} \left( y_{\frac{N}{2}+i} - y_{i} + EA_{m} - EA_{m} \right)^{2}  \nonumber \\
        &=& \frac{1}{N-1}\sum_{i=1}^{N/2}\left[ \left(y_{\frac{N}{2}+i}  - EA_{m} \right) - \left( y_{i} - EA_{m}\right) \right]^{2}  \nonumber  \\
        &=& \frac{1}{N-1}\sum_{i=1}^{N/2} \biggl[\left(y_{\frac{N}{2}+i}  - EA_{m}  \right)^2  + \left( y_{i} - EA_{m}\right)^2  \nonumber \\
        & &  - 2\left(y_{\frac{N}{2}+i}  - EA_{m}\right)  \times \left( y_{i} - EA_{m}\right) \biggr] \nonumber \\
        &=&  ED^{2}_{\sigma_m} - 2\times {\rm Cov}(y_{N-i},y_{i})
 \label{eq_proofstd2}        
\end{eqnarray}

\noindent where $Cov$ denotes covariance. Indeed, the second term in these 
equations is additive since the covariance among $Y^{-}$ and $Y^{+}$ is
negative.

Asymptotically the identities given by 
Eqs~\ref{proofabsstd}, \ref{eq_proofstd1}, and \ref{eq_proofstd2} are also
valid for odd numbers of measurements since $ED_{m} \simeq D_{m}$ and 
$ED_{\sigma_m} \simeq D_{\sigma_m}$. Similar identities that link even-dispersion 
parameters with their correspondents can be found using $A_{\mu}$, $A_{m}$, $EA_{\mu}$, and $EA_{m}$.

The dispersion of a distribution given by Eqs.~\ref{eq_defd1} and 
\ref{eq_defd2} is the standard deviation about the averages minus two times the
covariance among $Y^{-}$ and $Y^{+}$. Moreover, for symmetric distributions 
where $y_{N-i} - EA_{M} = -  \left( y_{i} - EA_{m}\right)$ and $EA_{m} =
EA_{\mu}$, we can write the following identity,

\begin{eqnarray}
 ED_{(1)} &=&   \sqrt{ED^{2}_{\sigma_m} - \frac{2}{N-1}\sum_{i=1}^{N/2} \left(y_{N-i}  - EA_{m}\right)  \times \left( y_{i} - EA_{m}\right)}  \nonumber \\
          &=&   \sqrt{ED^{2}_{\sigma_m} + \frac{2}{N-1}\sum_{i=1}^{N/2} \left( y_{i} - EA_{m}\right)^2}  \nonumber \\
          &=&  \sqrt{2}\times ED_{\sigma_m} 
 \label{eq_proofstd3}        
\end{eqnarray}

The ratio of $ED_{(1)}$ by $ED_{\sigma_m}$ can be used to estimate  whether
the measurements are symmetrically distributed.

\subsection{Shape parameters (S)}\label{sec_Shape_Parameters}

In a similar fashion as the dispersion parameters, we also can improve the
accuracy of skewness ($S_S$) and kurtosis ($S_K$) using the even-averages. 
Therefore, we propose even-skewness ($ES$) and even-kurtosis ($EK$) to
estimate the distribution shape (see lines 11-12 of Table~\ref{tb_varind_new}).
Moreover, we also propose higher moments of $ED_{(1)}$ and $ED_{(2)}$ as 
new even-shape parameters given by

\begin{equation}
 ES_{(1)} =   \frac{ \frac{1}{N} \sum_{i=1}^{N/2} \left(y_{N-i} - y_{i}\right)^{3} }{  ED_{(1)}^{3}  },
 \label{eq_defs1}        
\end{equation}

\begin{equation}
 ES_{(2)} =   \frac{ \frac{1}{N} \sum_{i=1}^{N/2} \left(y_{\frac{N}{2}+i} - y_{i}\right)^{3} }{  ED_{(2)}^{3}  },
 \label{eq_defs2}        
\end{equation}

\begin{equation}
 EK_{(1)} =   \frac{ \frac{1}{N} \sum_{i=1}^{N/2} \left(y_{N-i} - y_{i}\right)^{4} }{ ED_{(1)}^{4}  },
 \label{eq_defs3}        
\end{equation}

\noindent and  
 
\begin{equation}
 EK_{(2)} =   \frac{ \frac{1}{N} \sum_{i=1}^{N/2} \left(y_{\frac{N}{2}+i} - y_{i}\right)^{4} }{  ED_{(2)}^{4}  },
 \label{eq_defs4}        
\end{equation}

\noindent where $ES_{(1-2)}$ and $EK_{(1-2)}$ are unbound parameters, i.e. they are 
independent of the average. The values $ES_{(1-2)}$ mimic the skewness
while $EK_{(1-2)}$ mimic kurtosis. A strict relationship between $ES_{(1-2)}$ with
$ES$ and $EK_{(1-2)}$ with $EK$ is very complicated since such
definitions use distinct dispersion parameters. Indeed, we can use other dispersion 
parameters to broaden the list of even-shape parameters.

\subsection{Excess shape}\label{sec_Shape_Excess}

The integration of the Gaussian distribution returns a
kurtosis equal to 3 as $N \rightarrow \infty$.
An adjusted version of Pearson's kurtosis, the excess kurtosis, which is the
kurtosis minus 3 is most commonly used. Some authors refer to the excess
kurtosis as simply the "kurtosis". For example, the kurtosis function in IDL
language is actually the excess kurtosis. The excess values for even-shape
parameters were determined in the same fashion as the excess kurtosis, i.e.
shift these values to zero for normal distributions. Therefore, $10^5$ Monte Carlo 
simulations using the normal distributions with $10^6$ measurements were 
performed to determine the excess shape.

\begin{table}[htbp]
\addtolength{\tabcolsep}{3pt}  
 
\caption[]{Excess shape coefficients for even-shape
parameters.}\label{tab_excess}
 \centering 
 \begin{tabular}{| c | c | c |}        
  \hline
  $ES = 0.000(5)$  & $ES_{(1)} = 2.256(1)$ & $ES_{(2)} = 1.5190(6)$   \\[4pt]
  $EK = 3.000(2)$  & $EK_{(1)} = 6.00(1)$ & $EK_{(2)} = 2.461(3)$   \\[1pt]
  \hline
\end{tabular}
\end{table}

Table \ref{tab_excess} shows the averages for the even-shape parameters
and their errors. The amount of variation is less than $0.1\%$. The excess shape
values were added to the equations for even-shape parameters (see Table 
\ref{tb_varind_new}), rounding to two decimal places.

\section{Even interquartile range}\label{sec_eiqr}

The interquartile range \citet[IQR -][]{Kim-2014} is also included 
in our analysis since it was reported as one of the best statistical parameters
to select variable stars \citet[][]{Sokolovsky-2017}. The IQR uses the inner $50\%$
of measurements, excluding the $25\%$ brightest  and $25\%$ faintest flux
measurements, i.e. first the median value is computed in order to divide the 
set of measurements into upper and lower halves and then the IQR is given by 
the difference between the median values of the upper and lower halves.

The median value is improved using the even-median (see
Sect. \ref{sec_Accurate}) and hence the IQR can also be improved using the even
median instead of the median. Therefore, using the kernel defined by Eq.
\ref{eq_defvec} an even interquartile range is proposed as

\begin{equation}
 {\rm  EIQR} = EA_{m}(Y^{+}) - EA_{m}(Y^{-})
 \label{eq_eiqr}        
\end{equation}

\noindent where $Y^{+}$ and $Y^{-}$ are the measurements of
$Y$ above and below the median, respectively (see Sect. \ref{sec_Notation} for a
better description). Indeed, $EIQR$ provides adjustment for distributions with an
even or odd number of measurements. Three, two, or zero adjustments
are performed for distributions with odd, even (but not modulus 4), or modulus
4 numbers of measurements, respectively.

\section{Simulating distributions}\label{sec_Sumary_tests}

Monte Carlo simulations were used to test the even-statistical
parameters for a range of the number of measurements (number of epochs) varying
from $10$ to $100$. We performed $10^5$ simulations for a given number of measurements. 
Indeed, the range of measurements tested typifies light curves from 
current large wide-field, multi-epoch surveys such as Pan-STARRS, VVV, and Gaia. 
These simulations were performed for the five distributions described in  
Sect.~\ref{sec_Notation} that mimic noise and variable stars; the colours
and symbols in Fig.~\ref{fg_dist} are also adopted in the present section 
to facilitate the identification of these distributions. The statistical parameters have a
higher statistical significance for the distributions that have a large number of 
measurements, where the addition of measurements only leads to small
fluctuations.
Therefore, the adopted "true parameter'' values ($P_{true}$)
are those computed using $10^{6}$ measurements. We performed $10^{5}$ simulations 
to compute the $P_{true}$ values. All parameters analyzed are listed
in Table \ref{tab_mpar}; their errors were computed as the standard
deviation. This value is used as a reference to analyze the error given by

\begin{equation}
  e_{P} =   \frac{\left|P - P_{true}\right|}{P_{true}}
 \label{eq_error}        
\end{equation}

\noindent where $P$ means the statistical parameter. This expression provides
the mean error for $P$. In order to avoid singularities we shift the skewness
and kurtosis values for $P'_{true} = P_{true}+1$ and $P = P' + 1$ since 
they have $P_{true} \sim 0$.

\begin{figure*}[htb]
  \centering
  \includegraphics[width=0.49\textwidth,height=1.05\textwidth]{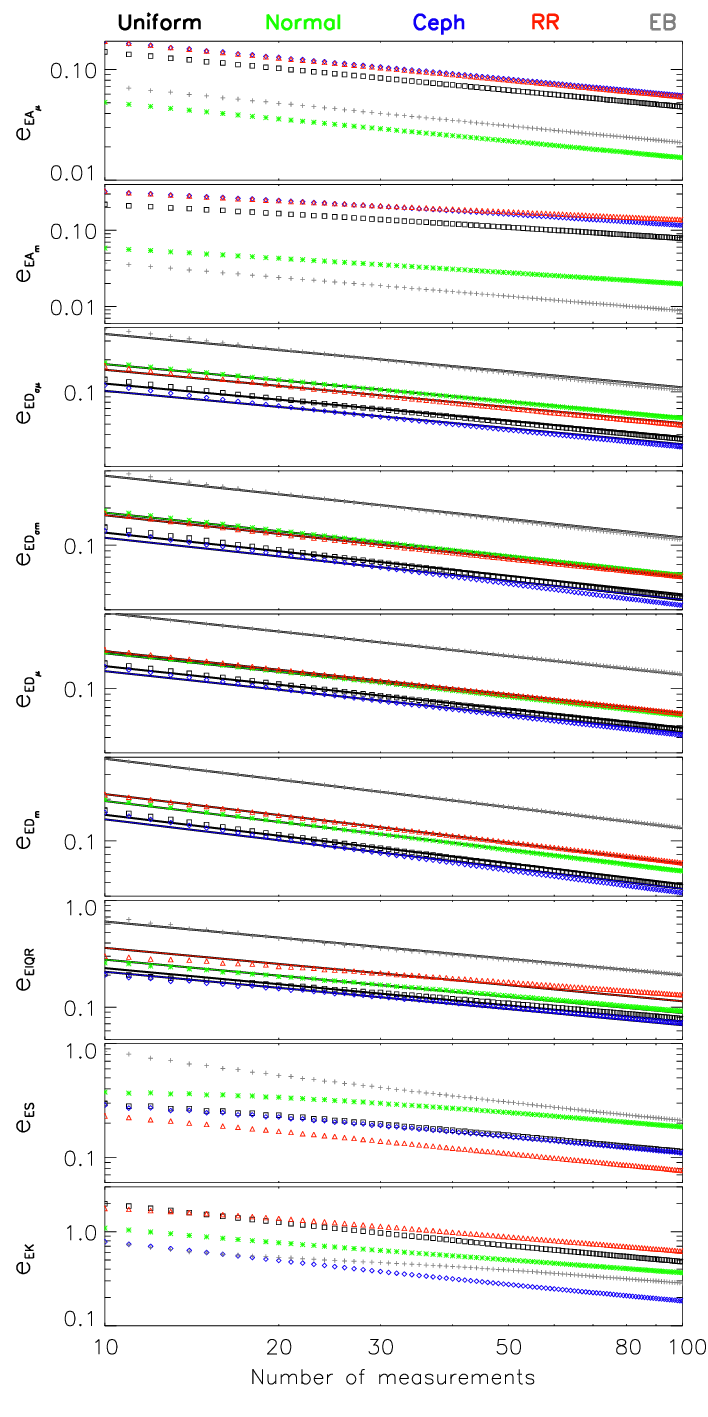}
  \includegraphics[width=0.49\textwidth,height=1.05\textwidth]{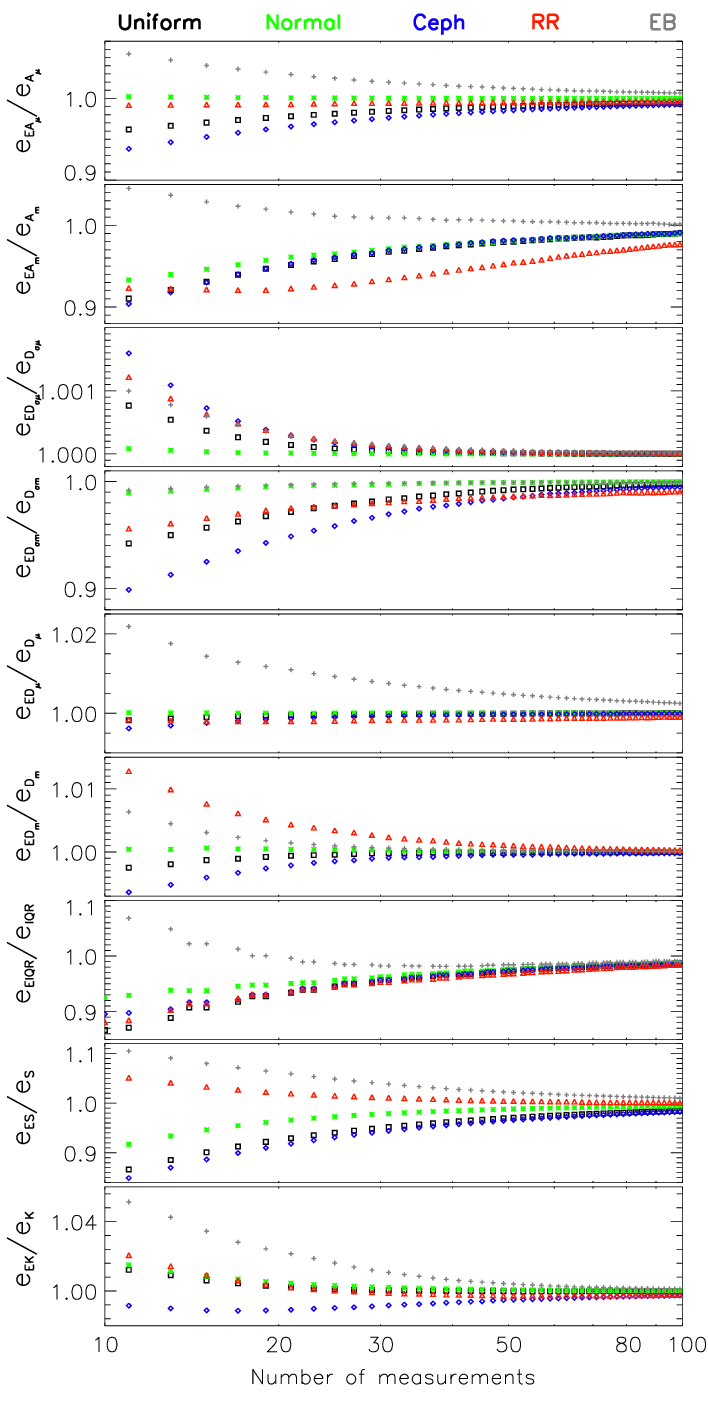}
  
  \caption{Parameter $e_{P}$ (left panels) and its comparison with previous statistical 
 parameters (right panels) as a function of the number of measurements (see 
 Tab.~\ref{tb_varind_new}). The colours and symbols are the same as those
 adopted in the Fig.~\ref{fg_dist}. In the left panels the results
 for the full range of measurements are used, while the right panel only shows
 those for odd numbers of measurements except for $\rm EIQR$, where 
 only the results with numbers of measurements that are not modulus 4 are
 plotted. The solid lines in the even-dispersion diagrams show the models
 described in Sect.~\ref{sec_weight}. }
 \label{fg_limmea1}
\end{figure*}

\begin{figure}[htb]
  \centering
  \includegraphics[width=0.49\textwidth,height=0.9\textwidth]{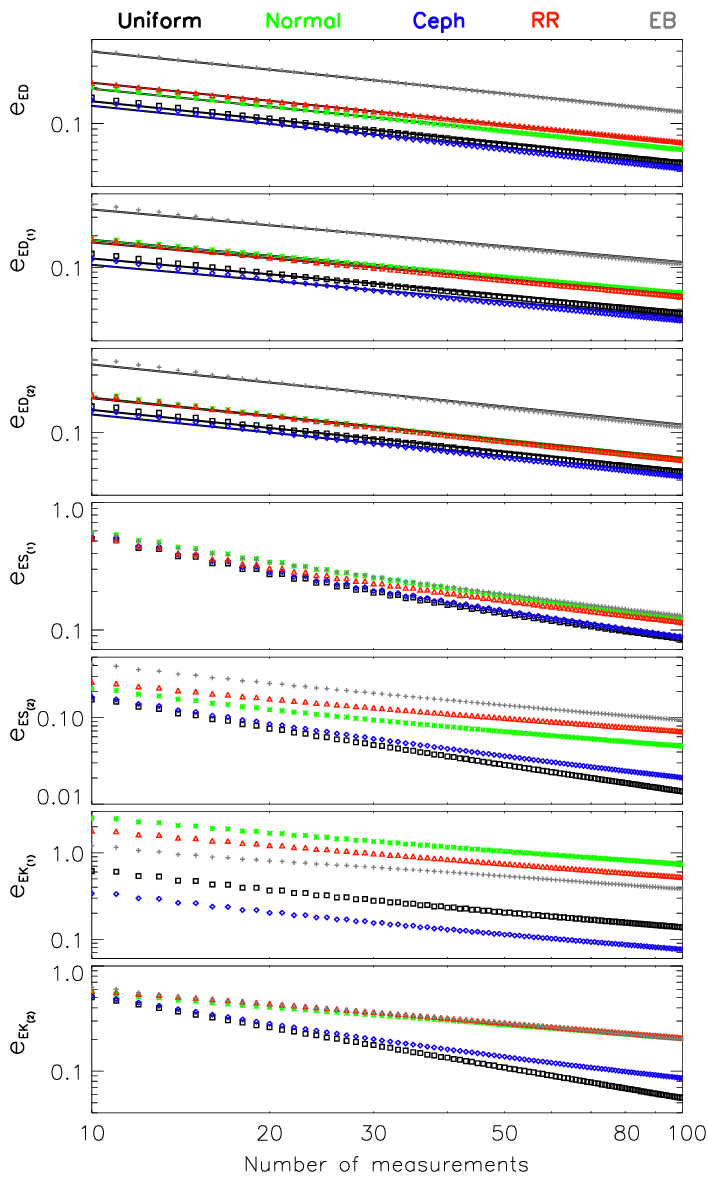}
  \caption{Parameter $e_{P}$ as a function of the number of measurements for the free 
  even-parameters (see Tabs.~\ref{tb_varind_new}). The colours are the same as in 
  Fig.~ \ref{fg_limmea1}.}
\label{fg_limmea2}
\end{figure}

\subsection{Bound even-statistical parameters}\label{sec_Accurate}

The bound even-statistical parameters are those dependent on the
average. These parameters only differ from the previous parameters in that the
mean and median are replaced by the even-mean and even-median. Figure~\ref{fg_limmea1} shows
$e_{P}$ (see Eq.\ref{eq_error}) for the even-parameters (see
Tab.~\ref{tb_varind_new} 1-6 and 10-12) and its 
comparison with previous parameters (mean, median, mean standard deviation, 
median standard deviation, mean absolute deviation, median absolute deviation, 
skewness, and kurtosis) as a function of the number of measurements in the left 
and right panels, respectively. The left panels include the results of
simulations for the whole range of measurements while the right panels only 
have the results for odd numbers of measurements because for even numbers of 
measurements the current and even-statistical parameters have the same values. 
Therefore, we only use the results for odd numbers of measurements, i.e. 
$11,13,15,\cdots$, where $e_{P}/e_{P'} < 1$ means a higher accuracy for the new
parameters compared to current parameters while $e_{P}/e_{P'} > 1$ means no 
improvement with the new parameters. The simulations were performed as described
in Sect.~\ref{sec_Sumary_tests}, from which we can observe that

\begin{itemize}

\item $EA_{\mu}$ $->$ The normal and EB distributions have similar
                  distributions, and separately Ceph, RR, and uniform
                  distributions are also similar.
                  The $EA_{\mu}$ returns the lowest errors for the normal
                  distribution.
                  Indeed, $e_{EA_{\mu}}/e_{A_{\mu}} \simeq 1$ for normal distributions while
                  $e_{EA_{\mu}}/e_{A_{\mu}} \simeq 1.07$ (for 10 measurements)
                  for the EB distribution. This happens because the dispersion
                  about the mean is symmetric for the normal distribution and
                  extremely asymmetric for the EB distribution.
                  The error for $10$ measurements for all distributions is about twice that found for 
                  $100$ measurements on average. The even-mean parameters are
                  more accurate than the mean for all distributions except the EB distribution. For instance,
                  the even-mean returns an improved accuracy of between $\sim
                  4\%$ and $\sim 8\%$ for Ceph, RR, and uniform distributions
                  over that found by the mean. On the other hand, the mean is
                  better then the even-mean by a similar rate for the EB
                  distribution.
\item  $EA_{m}$ $->$ The $e_{EA_{m}}$ for normal and EB distributions are similar
                  but offset by roughly a multiplicative factor.
                  The EB distribution $e_{EA_{m}}$ values are about twice
                  those found for the normal distribution for the same reason
                  as discussed for $e_{EA_{\mu}}$. The even-mean for the RR
                  distribution has $e_{EA_{\mu}}/e_{A_{\mu}} < 0.87$ for
                  $N<30$ measurements, i.e. an increase in the accuracy of about
                  $\sim 7\%$. Indeed, $EA_{m}$ along with $EA_{\mu}$ are more accurate than their previous definitions for the whole range of measurements and for four out of five distributions analyzed.

\item $ED_{\sigma_\mu}$ $->$ The EB distribution has the highest
                  $e_{ED_{\sigma_\mu}}$ values whereas the uniform,
                  Ceph, and RR distributions show similar values.
                  The same behaviour is also observed in the  $e_{ED_{\sigma_m}}$ diagram.
                  The $D_{\sigma_\mu}$ is more accurate than $ED_{\sigma_\mu}$ for the whole
                  range of measurements and distributions analyzed despite the improvement on the
                  estimation of the mean, but the difference is less than $\sim
                  0.2\%$ for Ceph distribution and less than $\sim 0.1\%$ otherwise.

\item $ED_{\sigma_m}$ $->$ This value is more accurate than $D_{\sigma_m}$ for
                  the whole range of measurements and all distributions analyzed.
                  The normal and EB distributions show
                  $e_{ED_{\sigma_m}}/e_{D_{\sigma_m}} \simeq 1$ for all range of
                  measurements analyzed. On the other hand, the Ceph and EB
                  distributions show
                  $e_{ED_{\sigma_m}}/e_{D_{\sigma_m}} \simeq 0.95$ for fewer
                  than $30$ measurements. The Ceph distribution has the lowest
                  value of $e_{ED_{\sigma_m}}/e_{D_{\sigma_m}}$, the opposite of
                  that found in $ED_{\sigma_\mu}$.
                  
\item  $ED_{\mu}$ and $ED_{m}$ $->$ The $e_{ED_{\mu}}$ and $e_{ED_{m}}$
                  are  similar to the EB distribution returning the highest values, while the other distributions return similar values to each other.
                  The $e_{ED_{\mu}}/e_{D_{\mu}}$ shows values less than one for all
                  distributions unless EB distribution. On the other hand,
                  $e_{ED_{m}}/e_{D_{m}}$ shows values of less than one for uniform
                  and Ceph distributions, about one for the normal distribution,
                  and greater than one for RR and EB distributions. The
                  largest value for $e_{ED_{\mu}}/e_{D_{\mu}} \simeq 1.005$ and
                  $e_{ED_{m}}/e_{D_{m}} \simeq 1.023$.

\item  ${\rm EIQR}$ $->$ The greatest variation on the
                  accuracy of statistical dispersion parameters is found for
                  ${\rm EIQR}$. An improvement of about $10\%$ is found for
                  uniform, normal, RR, and Ceph distributions and in opposite way a
                  diminishment in the accuracy of $10\%$ is found for the EB
                  distribution. 
                  All distributions with more than $\sim 30$ measurements show an 
                  improvement or similar accuracy as those found for IQR.  
                  The repeating patterns found among each consecutive set of
                  three measurements is related to the number of adjustments 
                  performed by the even-median (see Sect. \ref{sec_eiqr}).

\item $ES$ and $EK$ $->$ The shape parameters have the highest 
                  uncertainties among the statistical parameters analyzed.
                  The lowest values for $e_{ES}$ and highest values for $e_{EK}$
                  are found for the RR distribution, respectively.
                  The $e_{ES}/e_{S} < 1$ for all distributions except EB and RR
                  distributions while  $e_{EK}/e_{K} > 1$  for
                  all distributions except the Ceph distribution. Moreover
                  $e_{ES}$ and $e_{EK}$ have an uncertainty greater than $\sim
                  10\%$ up to $50$ measurements.
 
\end{itemize}

To summarize, the accuracy of statistical parameters has a strong 
dependence on the number of measurements and distribution type.
The even-mean and median are more accurate than the mean and median for all distributions analyzed except for the EB distribution.
The improvements in estimation of averages by even-statistics allows us to 
improve the estimation of dispersion and shape parameters for many 
distributions analyzed. It is mainly observed for distributions where the 
probability of finding measurements near $P_{true}$ is lower. As a result,  
$e_{P_{even}}/e_{P} \simeq 1$ for the normal distribution.
The even-statistical parameters are strongly dependent on the distribution shape and so they can be useful for discriminating distribution types. Therefore a study about how to classify distributions using even-statistical parameters will be performed in a later paper from this project.

\begin{table}[htbp]
 \scriptsize 
\caption[]{$P_{true}$ for all statistical parameters analyzed in the present work.}\label{tab_mpar}    
 \centering 
 \begin{tabular}{| c |  c |  c | c | c | c |}        
 \hline                
 $P_{true}$  & Uniform & Normal & Ceph & RR & EB \\
 \hline 

$A_{\mu}$ & 0.5000(3) & 0.5000(1) & 0.4743(3) & 0.3827(3) & 0.8412(2) \\ 
$EA_{\mu}$ & 0.5000(3) & 0.5000(1) & 0.4743(3) & 0.3827(3) & 0.8412(2) \\ 
$A_{m}$ & 0.5000(5) & 0.5000(1) & 0.4609(7) & 0.2798(6) & 0.93456(8) \\ 
$EA_{m}$ & 0.5000(5) & 0.5000(1) & 0.4609(7) & 0.2798(6) & 0.93456(8) \\ 
$D_{\sigma_\mu}$ & 0.2887(1) & 0.10000(7) & 0.3464(1) & 0.2726(2) & 0.2291(3) \\ 
$ED_{\sigma_\mu}$ & 0.2887(1) & 0.10000(7) & 0.3464(1) & 0.2726(2) & 0.2291(3) \\ 
$D_{\sigma_m}$ & 0.2887(1) & 0.10000(7) & 0.3467(1) & 0.2914(2) & 0.2474(3) \\ 
$ED_{\sigma_m}$ & 0.2887(1) & 0.10000(7) & 0.3467(1) & 0.2914(2) & 0.2474(3) \\ 
$D_{\mu}$ & 0.2500(1) & 0.07979(6) & 0.3084(2) & 0.2301(2) & 0.1579(3) \\ 
$ED_{\mu}$ & 0.2500(1) & 0.07979(6) & 0.3084(2) & 0.2301(2) & 0.1579(3) \\ 
$D_{m}$ & 0.2500(1) & 0.07979(6) & 0.3083(2) & 0.2199(2) & 0.1336(2) \\ 
$ED_{m}$ & 0.2500(1) & 0.07979(6) & 0.3083(2) & 0.2199(2) & 0.1336(2) \\ 
$S$ & 0.000(1) & 0.000(2) & 0.105(2) & 0.785(2) & -2.252(3) \\
$ES$ & 0.000(1) & 0.000(2) & 0.105(2) & 0.785(2) & -2.252(3) \\ 
$K$ & -1.200(1) & 0.000(5) & -1.4370(9) & -0.525(4) & 4.54(2) \\ 
$EK$ & -1.200(1) & 0.000(5) & -1.4370(9) & -0.525(4) & 4.54(2) \\ 
$ED$ & 0.5000(3) & 0.1596(1) & 0.6165(3) & 0.4397(4) & 0.2671(4) \\ 
$ED_{(1)}$ & 0.4082(2) & 0.14142(1) & 0.4893(2) & 0.3696(2) & 0.2723(4) \\ 
$ED_{(2)}$ & 0.3536(2) & 0.11511(8) & 0.4382(2) & 0.3308(2) & 0.2559(3) \\ 
$ES_{(1)}$ & -0.4227(5) & -0.003(1) & -0.5345(4) & -0.280(1) & 0.460(2) \\ 
$ES_{(2)}$ & -0.1056(3) & -0.0008(6) & -0.0847(2) & 0.152(1) & 1.145(2) \\ 
$EK_{(1)}$ & -2.400(2) & 0.00(1) & -2.895(2) & -1.717(5) & 2.34(2) \\ 
$EK_{(2)}$ & -0.4598(6) & 0.000(3) & -0.3827(6) & 0.564(4) & 5.67(2) \\ 

\hline 
\end{tabular}
\end{table}

\subsection{Unbound even-statistical parameters}\label{sec_unbound}

Unbound even-statistical parameters keep some relations with their
counterparts for particular limits and distributions (see
Sect.~\ref{sec_Dispersion_Parameters}). Such relations are not valid in general 
and therefore did not have a counterpart for comparison. Monte Carlo simulations 
were used to estimate the relative error of unbound even-statistical parameters 
such as those performed in Sect.~\ref{sec_Accurate}.

The unbound even-statistical parameters display similar $e_{P}$ values
as those found for previous parameters (see Fig.~\ref{fg_limmea2}); in those cases, bound 
(see Sect.~\ref{sec_Accurate}) and unbound even-dispersion parameters show
similar $e_{P}$ values, while the even-shape parameters return smaller errors
than the skewness and kurtosis. It means that the errors are comparable with
those found for the common statistical parameters used to describe
distributions.
Moreover, the even-statistical parameters are various for the different
distributions analyzed and so these parameters can be used to discriminate between such distributions (see
Table~\ref{tab_mpar}). Of course these values are for the distributions
described in Sect.~\ref{sec_Notation} that were generated to have the
same amplitude. The values are different if the amplitude is modified, for
instance.

The bound and unbound even-statistical parameters (see
Table~\ref{tb_varind_new}) have a similar accuracy to previous statistical 
parameters and hence they can be used to characterize statistical distributions in 
a similar fashion to previous distributions. Such parameters can be used to describe and 
differentiate distribution types. A better investigation about how use these parameters to describe various  
distributions will be performed in a forthcoming paper of
this project.

\subsection{Coefficients adjusted for sample size}\label{sec_weight}

The adjusted coefficients for sample size are used because samples
with few measurements have larger fluctuations in the estimated parameters.
For instance, the Fisher-Pearson coefficient (given by $\sqrt{n\times(n-1)}/(n-1)$) 
for a sample with $10$ and $100$ measurements is $1.054$ and $1.005$, 
respectively. As result, for instance, this correction increases the value if
the skewness is positive, and makes the value more negative if the skewness is 
negative. It cannot be used for parameters that only assume positive values 
such as standard deviation. Therefore other adjusted coefficients have been proposed 
in a similar fashion. These coefficients increase the dispersion in
a population since they enlarge the range of values. 

A single equation to create coefficients to adjust for sample size has been used
for all statistical parameter. However, the best adjustment is found using a
specific equation for each statistical parameters, since they each have different accuracies (see
Sect.~\ref{sec_Sumary_tests}).
The simulations described in Sect.~\ref{sec_Sumary_tests} were used to 
determine a model for each dispersion statistical parameter given by

\begin{equation}
  w_{(P)} =   1 - b_{(P)}\times N^{-1/2}, 
 \label{eq_coefnum}        
\end{equation}

\noindent where $b_{(P)}$ is a real number constant (see 
Table~\ref{tab_coef}). For unknown distributions, i.e. not included in our
analysis, the even-mean value may be used.

\begin{table}[htbp]
\caption[]{Coefficients (b) of Eq.\ref{eq_coefnum}; the last column (All) is the even-mean of the values found for all 
distributions analyzed.}\label{tab_coef}
 \scriptsize   
 \centering 
 \begin{tabular}{| c | c | c | c | c | c |  c |}        
 \hline                
 \multicolumn{1}{|c}{ $b_{(P)}$}  &  \multicolumn{1}{|c}{Uniform}  & \multicolumn{1}{|c|}{Normal}  & \multicolumn{1}{|c|}{Ceph}  & \multicolumn{1}{|c|}{RR}  & \multicolumn{1}{|c|}{EB}   & \multicolumn{1}{|c|}{All}  \\ 
 \hline 
$ED_{\sigma_\mu}$ & $0.379$  & $0.575$  & $0.324$  & $0.510$  & $1.104$  & $0.477$  \\
$D_{\sigma_\mu}$ & $0.379$  & $0.575$  & $0.324$  & $0.509$  & $1.104$  & $0.477$  \\
$ED_{\sigma_m}$ & $0.399$  & $0.580$  & $0.363$  & $0.552$  & $1.156$  & $0.490$  \\
$D_{\sigma_m}$ & $0.410$  & $0.583$  & $0.376$  & $0.556$  & $1.161$  & $0.497$  \\
$ED_{\mu}$ & $0.479$  & $0.609$  & $0.436$  & $0.633$  & $1.292$  & $0.544$  \\
$D_{\mu}$ & $0.479$  & $0.609$  & $0.436$  & $0.634$  & $1.286$  & $0.544$  \\
$ED_{m}$ & $0.487$  & $0.612$  & $0.452$  & $0.686$  & $1.235$  & $0.550$  \\
$D_{m}$ & $0.487$  & $0.612$  & $0.452$  & $0.685$  & $1.234$  & $0.550$  \\
EIQR & $0.735$  & $0.883$  & $0.679$  & $1.137$  & $2.007$  & $0.809$  \\
IQR & $0.770$  & $0.912$  & $0.708$  & $1.189$  & $2.000$  & $0.841$  \\
$ED$ & $0.484$  & $0.616$  & $0.444$  & $0.691$  & $1.254$  & $0.550$  \\
$ED_{(1)}$ & $0.384$  & $0.578$  & $0.334$  & $0.546$  & $1.124$  & $0.481$  \\
$ED_{(2)}$ & $0.486$  & $0.614$  & $0.446$  & $0.608$  & $1.160$  & $0.550$  \\

\hline 
\end{tabular}
\end{table}

\section{Modelling the noise}\label{sec_noise_model}

\citet[][]{Cross-2009} used the Strateva function \citep[see][for more 
details]{Strateva-2001,Sesar-2007} to fit the standard deviation as a function
of magnitude to estimate a noise model ($\zeta$). This method assumes that the 
majority of the sample are point sources, where the variability measurements are
dominated by noise, rather than astrophysical variations. The method provides a suitable
model for photometric surveys at optical wavelengths if they have a single
component of noise that increases in relative magnitude from bright to faint
stars. However, the brightest stars can show much greater variation which comes
from saturation and non-linearity of the detectors providing a source of
variation that cannot be fit by these models. Such a situation is rare at optical
wavebands but is very frequently present for NIR data (see Fig.
\ref{fg_magdispshape}). The sky foreground emitted by the atmosphere is 
highly variable in the NIR. For this reason, the sky foreground causes a highly time-varying saturation limit, 
which can affect large parts of otherwise highly accurate time-series data for 
bright stars with substantial outliers that have very small formal error estimates 
\citep[][]{FerreiraLopes-2015wfcam}. These outliers probably lead to a 
spurious impact upon the statistical parameters. Therefore, we propose a 
modification to the Strateva function that allows us to model such variations,
the increase in the standard deviation for bright (saturated stars) and faint
(photon noise) stars, given by

\begin{equation}
  \zeta_{(P)} ( m ) = c_{0} + c_{1} 10^{0.4\times m} + c_{2} 10^{0.5\times m}+ c_{3} 10^{-1.4\times m},
 \label{eq_stratevamod}        
\end{equation}

\noindent  where all the coefficients are real numbers. Indeed, \citet[][]{Strateva-2001} and \citet[][]{Sesar-2007} proposed a noise model using three terms where the second and third coefficients are $0.4$ and $0.8$, respectively. 
These powers continued to be used in \citet[][]{Cross-2009} but the optimal
coefficients were never tested. For more details see Sect. \ref{sec_test_noisemodel}.

\subsection{Non-correlated indices}\label{subsec_index}

The selection of non-stochastic variations can be performed by one or more dispersion parameters.
In order to combine a set of dispersion parameters (see Table
\ref{tb_varind_new}) a non-correlated index is proposed as follows;

\begin{equation}
  I_{(P)} = \frac{ w_{(P)} \times P }{\zeta_{w_{(P)}P} } 
 \label{eq_varind1}        
\end{equation}

\noindent where $w_{(P)}$ and $\zeta_{w_{(P)}P}$ are given by Eqs.
\ref{eq_coefnum} and \ref{eq_stratevamod}, respectively. 
This equation provides an index value that takes into the sample
size adjustment coefficient and noise model into account. For instance, $I_{(P)} \sim 1$
for stochastic variation. 

Distinct statistical parameters have different capabilities of
discriminating distributions (see Sects. \ref{sec_Accurate} and
\ref{sec_unbound}). Such differences are highlighted when the $e_{P}$ values or
sample size adjustment coefficients are compared. A sample composed mainly of
stochastic variations has a different dispersion of $I_{(P)}$ values. 
Therefore an appropriate combination of the results from different dispersion parameters is given by

\begin{equation}
  X_{f} = \frac{ \sum_{j=1}^{v} \omega_{P_{j}} I_{P_{j}} }{\sum_{j=1}^{v} \omega_{P_{j}}},
 \label{eq_varind2}        
\end{equation}

\noindent where $f$ is the waveband used, $\omega_{P_{j}}$ is a weight related 
with each dispersion parameter, $v$ is the number of parameters 
used, and $I_{P_{j}}$ is given by Eq.~\ref{eq_varind1}. Indeed, $I_{P_{j}}$
provides a normalized index allowing us to combine distinct dispersion
parameters and the results from various wavebands. 

\subsection{Broadband selection}\label{subsec_cutoffs}

Variable stars candidates for non-correlated data are usually selected from the 
noise model (see Sect. \ref{sec_noise_model}). Stars with values above 
$n\times D_{\sigma\mu}$ are selected for further analyses. This approach 
assumes that a few percent of entire sample are variable stars and have
statistical values above the noise. The noise samples present distributions 
such as uniform, normal, or distributions in between, while variable
stars are more similar to the Ceph, RR, and EB distributions.
Therefore, the dispersion parameters assume a different range of values for variable and non-variable
stars that is mainly highlighted for samples with a high number of
measurements (typically higher than 50). Indeed, such a difference must increase
for higher amplitudes than that found for the noise. For few measurements 
(typically less than 20) stochastic and non-stochastic variations have large
uncertainties increasing the mis-selection rate (see
Sect.~\ref{sec_Accurate}). We find similar behaviour for correlated indices. 
For instance, \citet[][]{FerreiraLopes-2015wfcam} use cut-off surfaces linking
magnitude, number of epochs, and variability indices to improve the selection 
criteria of variable stars, while \citet[][]{FerreiraLopes-2016I} use flux 
independent indices to propose an empirical relationship between cut-off values
and the number of measurements without taking into account magnitude.

The adjusted coefficients for sample size, as presented in
Sect.~\ref{sec_weight}, reduce the population dispersion. Meanwhile,
uncertainties about the range of values assumed by stochastic and non-stochastic
variations also vary with the number of measurements. For non-stochastic 
variations with a good signal-to-noise and a large number of measurements, such
a range is different to that produced by stochastic variations. On the other
hand, for distributions with just a few measurements, the range of values can
significantly overlap. In the same fashion as the empirical selection criteria
proposed by \citealt[][]{FerreiraLopes-2016I} (see Eq. 16), we propose the
follow criteria,

\begin{equation}
  f(\alpha,\beta) = \alpha+\sqrt{\frac{\beta}{N}}
 \label{eq_bcutoff1}        
\end{equation}

\noindent where $\alpha$ and $\beta$ are real positive values and $N$ is
the number of measurements. In these case where $\alpha$ is bigger than 1, we may find 
stochastic variations. Higher values of $\beta$ provide a higher cut-off
for small numbers of measurements or correlations. For instance, $f(1,4)$ for $N$
equal to 10, 30, and 50 are $1.63$, $1.36$, and $1.28$, respectively. Indeed 
lower values of $\alpha$ provide a more complete selection while higher values 
provide a more reliable selection.

\section{Real data}\label{sec_real_data}

We used the  WFCAM Calibration 08B release \citep[WFCAMCAL08B - ][]{Hodgkin-2009,
Cross-2009} as a test database as we did in the first paper of this series.
To summarize, this programme contains panchromatic data for 58 different
pointings distributed over the full range in right ascension and spread over 
declinations of $+59\fdg62$ and $-24\fdg73$. These data have been used to 
calibrate the UKIDSS surveys \citep[][]{Lawrence-2007}. During each visit the 
fields were usually observed with a sequence of filters, either JHK
or ZYJHK within a few minutes. This led to an irregular sampling with 
fields reobserved roughly on a daily basis, although longer time gaps are
common and of course large seasonal gaps are also present in the data set. For 
more information about design, the details of the data curation procedures, the 
layout, and variability analysis on this database are described in detail
in \citealt[][]{Hambly-2008}, \citealt[][]{Cross-2009}, and 
\citealt[][]{FerreiraLopes-2015wfcam}.

\begin{figure*}[htb]
  \centering
  \includegraphics[width=0.49\textwidth,height=1.\textwidth]{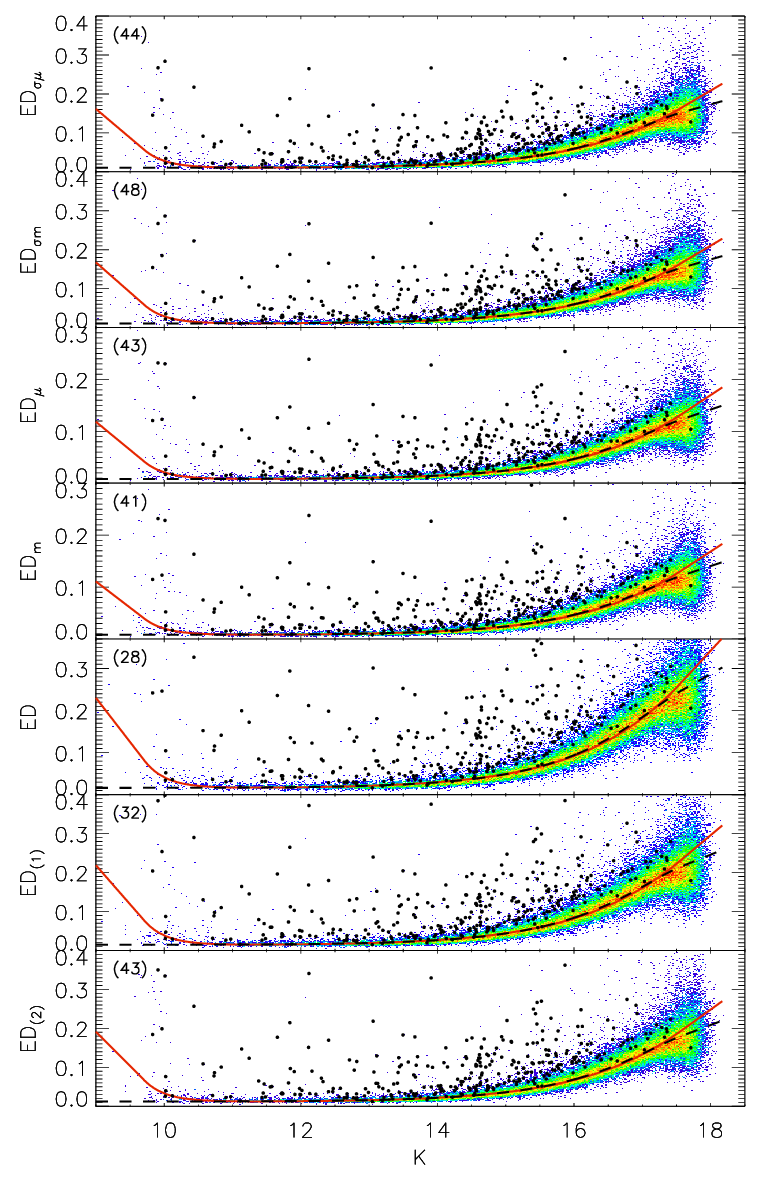}
  \includegraphics[width=0.49\textwidth,height=1.\textwidth]{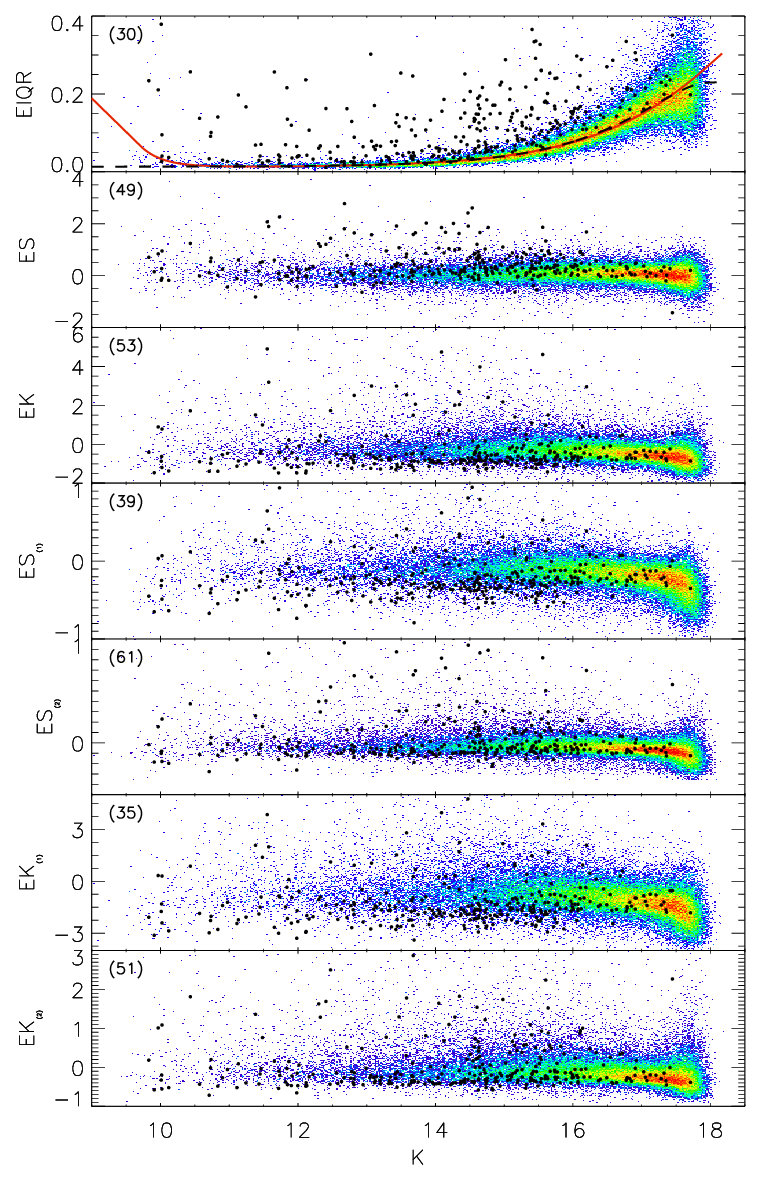}

  \caption{Dispersion and shape parameters as a function of magnitude where the 
  black dots indicate the WVSC1 stars. The red and dashed black lines indicate the 
  Strateva-modified and Strateva functions, respectively. The maximum
  number of sources per pixel is shown in brackets in each panel. }
 \label{fg_magdispshape}
\end{figure*}

The multi-waveband data were well fitted to test the statistical parameters
using different wavebands ($ZYJHK$). Moreover, \citealt[][]{FerreiraLopes-2015wfcam}
and \citealt[][]{FerreiraLopes-2016I} performed a comprehensive stellar 
variability analysis of the WFCAMCAL08B characterizing the photometric data and
identifying $319$ stars (WVSC1), of which $275$ are classified as periodic
variable stars and $44$ objects as suspected variables or apparent aperiodic 
variables. In this paper we analyze the same sample from 
\citealt[][]{FerreiraLopes-2015wfcam} and \citealt[][]{FerreiraLopes-2016I}. First, 
we selected all sources classified as a star or probable star with at least
10 unflagged epochs in any of the five filters. This selection was performed 
from an initial database of $216,722$ stars. Next we test the efficiency of 
selection of variable stars using the statistical parameters presented in 
Sect.~\ref{sec_Even_Statistic}.

We compute all statistical parameters listed in Table~\ref{tb_varind_new}
by the following algorithm: the photometry measured by the best aperture was
selected and next the measurements with flags (\verb+ppErrBits+) higher than 256
were removed. The analysis of these data was performed
using the current and earlier approaches, where the comparison between the
current and earlier approaches was tested using the following equation:
 
\begin{equation}
  G(P) = 100\times\left(\frac{P' - P}{P'}\right)
 \label{eq_estimate}        
\end{equation}

\noindent  where $G(P)$ means the percentage of upgrade ($G
> 0$) or downgrade ($G < 0$) provided by the parameter tested ($P$). For
example, $P = E_{tot}$ (ratio of the total number of sources selected to the 
total number of variable stars in the WVSC1 catalogue) and  $P = E_{WVSC1}$
(ratio of number of selected variables stars in WVSC1 to the total number of 
variable stars in WVSC1) computed from previous ($P'$) and current ($P$)
statistics.
This allows us estimate the improvement ($G > 0$) or deterioration ($G < 0$)
provided by the methods proposed in the current work (see Sect. 
\ref{sec_testupgrade}). The statistic parameters were computed for each waveband as well as
considering all wavebands ($ZYJHK$). Table~\ref{tab_ef} lists $\alpha$ and
its respective efficiency metric values. Such parameters are used to analyze the
efficiency of selection of variable stars from noisy data in the WFCAMCAL
database using the WVSC1 catalogue as a comparison.

\subsection{Testing even-statistical parameters}\label{sec_testeven}

Figure \ref{fg_magdispshape} shows even-statistical parameters and the standard
deviation as a function of the K-band magnitude. The variable stars in WVSC1 are
denoted by large black dots and the noise model (Strateva)
functions are indicated by lines. The main results can be summarized as follows:

\begin{table*}[htbp]
 \scriptsize 
\caption[]{Strateva and Strateva modified parameters (see 
Eq. \ref{eq_stratevamod}) for all dispersion parameters analyzed in the present 
work for $BA$. The metric G to measure upgrade ($G > 0$) or downgrade ($G < 0$) 
for $\chi^2$ and dispersion of the residuals (R) is also listed in each
line.}\label{tab_noisemodel}
 \centering 
 \begin{tabular}{| c | c | c  c  c  | c c c c | c c |}        
 \hline                

 \multicolumn{1}{|c}{}  & \multicolumn{1}{|c}{}  &  \multicolumn{3}{|c}{ Strateva }  & \multicolumn{4}{|c}{ Strateva Modified } &  \multicolumn{2}{|c|}{  }  \\ 
& & $c_{0}$ & $c_{1}$ & $c_{2}$ & $c_{0}$ & $c_{1}$ & $c_{2}$ & $c_{3}$ & $G(\, \chi^2 \,)$ & $G(\, R \,)$ \\ 
 \hline
&Z&$2.447\times10^{-2}$&$3.243\times10^{-9}$&$-1.459\times10^{-17}$&$2.175\times10^{-2}$&$6.603\times10^{-9}$&$-4.836\times10^{-11}$&$4.392\times10^{12}$&$83.0\%$&$0.8\%$\\
&Y&$1.633\times10^{-2}$&$4.115\times10^{-9}$&$-2.258\times10^{-17}$&$1.284\times10^{-2}$&$8.479\times10^{-9}$&$-6.555\times10^{-11}$&$1.323\times10^{13}$&$74.6\%$&$1.1\%$\\
$ED_{\sigma_\mu}$&J&$1.978\times10^{-2}$&$8.899\times10^{-9}$&$-1.154\times10^{-16}$&$1.300\times10^{-2}$&$1.962\times10^{-8}$&$-1.913\times10^{-10}$&$7.196\times10^{12}$&$88.5\%$&$1.3\%$\\
&H&$2.052\times10^{-2}$&$1.400\times10^{-8}$&$-2.746\times10^{-16}$&$9.926\times10^{-3}$&$3.090\times10^{-8}$&$-3.287\times10^{-10}$&$2.207\times10^{13}$&$95.4\%$&$1.1\%$\\
&K&$2.446\times10^{-2}$&$2.326\times10^{-8}$&$-8.375\times10^{-16}$&$1.160\times10^{-2}$&$5.403\times10^{-8}$&$-6.763\times10^{-10}$&$4.591\times10^{12}$&$95.8\%$&$1.2\%$\\
\hline
&Z&$2.447\times10^{-2}$&$3.243\times10^{-9}$&$-1.459\times10^{-17}$&$2.175\times10^{-2}$&$6.603\times10^{-9}$&$-4.836\times10^{-11}$&$4.392\times10^{12}$&$83.0\%$&$0.8\%$\\
&Y&$1.632\times10^{-2}$&$4.115\times10^{-9}$&$-2.258\times10^{-17}$&$1.284\times10^{-2}$&$8.479\times10^{-9}$&$-6.555\times10^{-11}$&$1.323\times10^{13}$&$74.6\%$&$1.1\%$\\
$D_{\sigma_\mu}$&J&$1.978\times10^{-2}$&$8.899\times10^{-9}$&$-1.154\times10^{-16}$&$1.300\times10^{-2}$&$1.962\times10^{-8}$&$-1.913\times10^{-10}$&$7.196\times10^{12}$&$88.5\%$&$1.3\%$\\
&H&$2.052\times10^{-2}$&$1.400\times10^{-8}$&$-2.746\times10^{-16}$&$9.926\times10^{-3}$&$3.090\times10^{-8}$&$-3.287\times10^{-10}$&$2.207\times10^{13}$&$95.4\%$&$1.1\%$\\
&K&$2.446\times10^{-2}$&$2.326\times10^{-8}$&$-8.375\times10^{-16}$&$1.160\times10^{-2}$&$5.403\times10^{-8}$&$-6.763\times10^{-10}$&$4.591\times10^{12}$&$95.8\%$&$1.2\%$\\
\hline
&Z&$2.469\times10^{-2}$&$3.258\times10^{-9}$&$-1.460\times10^{-17}$&$2.197\times10^{-2}$&$6.620\times10^{-9}$&$-4.841\times10^{-11}$&$4.445\times10^{12}$&$83.5\%$&$0.7\%$\\
&Y&$1.643\times10^{-2}$&$4.139\times10^{-9}$&$-2.268\times10^{-17}$&$1.289\times10^{-2}$&$8.522\times10^{-9}$&$-6.582\times10^{-11}$&$1.388\times10^{13}$&$76.0\%$&$1.0\%$\\
$ED_{\sigma_m}$&J&$1.993\times10^{-2}$&$8.913\times10^{-9}$&$-1.145\times10^{-16}$&$1.310\times10^{-2}$&$1.959\times10^{-8}$&$-1.904\times10^{-10}$&$7.296\times10^{12}$&$88.7\%$&$1.2\%$\\
&H&$2.092\times10^{-2}$&$1.393\times10^{-8}$&$-2.677\times10^{-16}$&$1.004\times10^{-2}$&$3.075\times10^{-8}$&$-3.252\times10^{-10}$&$2.286\times10^{13}$&$95.4\%$&$1.1\%$\\
&K&$2.482\times10^{-2}$&$2.320\times10^{-8}$&$-8.237\times10^{-16}$&$1.173\times10^{-2}$&$5.393\times10^{-8}$&$-6.725\times10^{-10}$&$4.694\times10^{12}$&$95.8\%$&$1.2\%$\\
\hline
&Z&$2.470\times10^{-2}$&$3.256\times10^{-9}$&$-1.456\times10^{-17}$&$2.198\times10^{-2}$&$6.612\times10^{-9}$&$-4.830\times10^{-11}$&$4.682\times10^{12}$&$83.3\%$&$0.7\%$\\
&Y&$1.642\times10^{-2}$&$4.138\times10^{-9}$&$-2.266\times10^{-17}$&$1.291\times10^{-2}$&$8.514\times10^{-9}$&$-6.573\times10^{-11}$&$1.365\times10^{13}$&$75.9\%$&$1.0\%$\\
$D_{\sigma_m}$&J&$1.994\times10^{-2}$&$8.913\times10^{-9}$&$-1.144\times10^{-16}$&$1.311\times10^{-2}$&$1.959\times10^{-8}$&$-1.902\times10^{-10}$&$7.306\times10^{12}$&$88.8\%$&$1.2\%$\\
&H&$2.089\times10^{-2}$&$1.395\times10^{-8}$&$-2.681\times10^{-16}$&$1.004\times10^{-2}$&$3.076\times10^{-8}$&$-3.251\times10^{-10}$&$2.280\times10^{13}$&$95.5\%$&$1.1\%$\\
&K&$2.483\times10^{-2}$&$2.320\times10^{-8}$&$-8.222\times10^{-16}$&$1.173\times10^{-2}$&$5.390\times10^{-8}$&$-6.718\times10^{-10}$&$4.700\times10^{12}$&$95.9\%$&$1.2\%$\\
\hline
&Z&$1.638\times10^{-2}$&$2.373\times10^{-9}$&$-1.007\times10^{-17}$&$1.405\times10^{-2}$&$4.718\times10^{-9}$&$-3.357\times10^{-11}$&$1.200\times10^{13}$&$62.6\%$&$0.6\%$\\
&Y&$1.180\times10^{-2}$&$2.983\times10^{-9}$&$-1.543\times10^{-17}$&$9.106\times10^{-3}$&$5.992\times10^{-9}$&$-4.501\times10^{-11}$&$1.331\times10^{13}$&$45.3\%$&$0.8\%$\\
$ED_{\mu}$&J&$1.312\times10^{-2}$&$6.674\times10^{-9}$&$-8.432\times10^{-17}$&$8.983\times10^{-3}$&$1.409\times10^{-8}$&$-1.346\times10^{-10}$&$4.139\times10^{12}$&$70.1\%$&$1.0\%$\\
&H&$1.408\times10^{-2}$&$1.056\times10^{-8}$&$-1.990\times10^{-16}$&$6.945\times10^{-3}$&$2.233\times10^{-8}$&$-2.313\times10^{-10}$&$1.484\times10^{13}$&$93.4\%$&$1.1\%$\\
&K&$1.532\times10^{-2}$&$1.832\times10^{-8}$&$-6.596\times10^{-16}$&$7.556\times10^{-3}$&$3.981\times10^{-8}$&$-4.888\times10^{-10}$&$2.688\times10^{12}$&$91.9\%$&$1.0\%$\\
\hline
&Z&$1.637\times10^{-2}$&$2.373\times10^{-9}$&$-1.008\times10^{-17}$&$1.405\times10^{-2}$&$4.719\times10^{-9}$&$-3.359\times10^{-11}$&$1.200\times10^{13}$&$62.5\%$&$0.6\%$\\
&Y&$1.179\times10^{-2}$&$2.983\times10^{-9}$&$-1.543\times10^{-17}$&$9.105\times10^{-3}$&$5.992\times10^{-9}$&$-4.501\times10^{-11}$&$1.330\times10^{13}$&$45.3\%$&$0.8\%$\\
$D_{\mu}$&J&$1.312\times10^{-2}$&$6.675\times10^{-9}$&$-8.438\times10^{-17}$&$8.979\times10^{-3}$&$1.409\times10^{-8}$&$-1.347\times10^{-10}$&$4.134\times10^{12}$&$69.9\%$&$1.0\%$\\
&H&$1.407\times10^{-2}$&$1.056\times10^{-8}$&$-1.991\times10^{-16}$&$6.940\times10^{-3}$&$2.234\times10^{-8}$&$-2.314\times10^{-10}$&$1.484\times10^{13}$&$93.4\%$&$1.1\%$\\
&K&$1.530\times10^{-2}$&$1.832\times10^{-8}$&$-6.604\times10^{-16}$&$7.555\times10^{-3}$&$3.982\times10^{-8}$&$-4.889\times10^{-10}$&$2.680\times10^{12}$&$91.9\%$&$1.0\%$\\
\hline
&Z&$1.605\times10^{-2}$&$2.365\times10^{-9}$&$-1.011\times10^{-17}$&$1.374\times10^{-2}$&$4.713\times10^{-9}$&$-3.365\times10^{-11}$&$1.170\times10^{13}$&$58.0\%$&$0.6\%$\\
&Y&$1.169\times10^{-2}$&$2.961\times10^{-9}$&$-1.536\times10^{-17}$&$9.025\times10^{-3}$&$5.955\times10^{-9}$&$-4.480\times10^{-11}$&$1.301\times10^{13}$&$43.0\%$&$0.8\%$\\
$ED_{m}$&J&$1.288\times10^{-2}$&$6.658\times10^{-9}$&$-8.472\times10^{-17}$&$8.894\times10^{-3}$&$1.405\times10^{-8}$&$-1.346\times10^{-10}$&$3.878\times10^{12}$&$65.2\%$&$1.1\%$\\
&H&$1.367\times10^{-2}$&$1.060\times10^{-8}$&$-2.036\times10^{-16}$&$6.831\times10^{-3}$&$2.237\times10^{-8}$&$-2.330\times10^{-10}$&$1.405\times10^{13}$&$92.2\%$&$1.0\%$\\
&K&$1.474\times10^{-2}$&$1.847\times10^{-8}$&$-6.796\times10^{-16}$&$7.455\times10^{-3}$&$3.987\times10^{-8}$&$-4.917\times10^{-10}$&$2.480\times10^{12}$&$89.8\%$&$0.8\%$\\
\hline
&Z&$1.605\times10^{-2}$&$2.365\times10^{-9}$&$-1.012\times10^{-17}$&$1.373\times10^{-2}$&$4.715\times10^{-9}$&$-3.368\times10^{-11}$&$1.171\times10^{13}$&$58.0\%$&$0.6\%$\\
&Y&$1.168\times10^{-2}$&$2.962\times10^{-9}$&$-1.540\times10^{-17}$&$9.017\times10^{-3}$&$5.961\times10^{-9}$&$-4.488\times10^{-11}$&$1.297\times10^{13}$&$42.6\%$&$0.9\%$\\
$D_{m}$&J&$1.288\times10^{-2}$&$6.657\times10^{-9}$&$-8.474\times10^{-17}$&$8.894\times10^{-3}$&$1.405\times10^{-8}$&$-1.346\times10^{-10}$&$3.871\times10^{12}$&$65.2\%$&$1.1\%$\\
&H&$1.367\times10^{-2}$&$1.060\times10^{-8}$&$-2.037\times10^{-16}$&$6.828\times10^{-3}$&$2.237\times10^{-8}$&$-2.331\times10^{-10}$&$1.405\times10^{13}$&$92.2\%$&$1.0\%$\\
&K&$1.473\times10^{-2}$&$1.848\times10^{-8}$&$-6.807\times10^{-16}$&$7.448\times10^{-3}$&$3.990\times10^{-8}$&$-4.923\times10^{-10}$&$2.479\times10^{12}$&$89.8\%$&$0.8\%$\\
\hline
&Z&$2.139\times10^{-2}$&$3.558\times10^{-9}$&$-1.384\times10^{-17}$&$1.824\times10^{-2}$&$6.532\times10^{-9}$&$-4.357\times10^{-11}$&$2.727\times10^{13}$&$-54.8\%$&$0.07\%$\\
&Y&$1.695\times10^{-2}$&$4.498\times10^{-9}$&$-2.223\times10^{-17}$&$1.327\times10^{-2}$&$8.563\times10^{-9}$&$-6.198\times10^{-11}$&$2.348\times10^{13}$&$-65.1\%$&$0.012\%$\\
EIQR&J&$1.787\times10^{-2}$&$9.982\times10^{-9}$&$-1.162\times10^{-16}$&$1.311\times10^{-2}$&$1.960\times10^{-8}$&$-1.781\times10^{-10}$&$4.595\times10^{12}$&$-4.1\%$&$0.005\%$\\
&H&$1.966\times10^{-2}$&$1.628\times10^{-8}$&$-2.867\times10^{-16}$&$9.761\times10^{-3}$&$3.263\times10^{-8}$&$-3.243\times10^{-10}$&$2.095\times10^{13}$&$88.1\%$&$0.5\%$\\
&K&$1.935\times10^{-2}$&$2.916\times10^{-8}$&$-9.982\times10^{-16}$&$9.832\times10^{-3}$&$5.897\times10^{-8}$&$-6.958\times10^{-10}$&$3.215\times10^{12}$&$81.5\%$&$0.3\%$\\
\hline
&Z&$2.128\times10^{-2}$&$3.565\times10^{-9}$&$-1.407\times10^{-17}$&$1.804\times10^{-2}$&$6.589\times10^{-9}$&$-4.429\times10^{-11}$&$2.833\times10^{13}$&$-57.0\%$&$0.05\%$\\
&Y&$1.690\times10^{-2}$&$4.502\times10^{-9}$&$-2.252\times10^{-17}$&$1.322\times10^{-2}$&$8.608\times10^{-9}$&$-6.266\times10^{-11}$&$2.309\times10^{13}$&$-73.6\%$&$0.005\%$\\
IQR&J&$1.774\times10^{-2}$&$1.002\times10^{-8}$&$-1.187\times10^{-16}$&$1.297\times10^{-2}$&$1.979\times10^{-8}$&$-1.812\times10^{-10}$&$4.555\times10^{12}$&$-12.2\%$&$0.02\%$\\
&H&$1.923\times10^{-2}$&$1.635\times10^{-8}$&$-2.917\times10^{-16}$&$9.812\times10^{-3}$&$3.263\times10^{-8}$&$-3.253\times10^{-10}$&$1.964\times10^{13}$&$85.8\%$&$0.4\%$\\
&K&$1.923\times10^{-2}$&$2.931\times10^{-8}$&$-1.024\times10^{-15}$&$9.769\times10^{-3}$&$5.952\times10^{-8}$&$-7.076\times10^{-10}$&$3.182\times10^{12}$&$80.8\%$&$0.18\%$\\
\hline
&Z&$3.206\times10^{-2}$&$4.802\times10^{-9}$&$-2.046\times10^{-17}$&$2.776\times10^{-2}$&$9.392\times10^{-9}$&$-6.648\times10^{-11}$&$2.318\times10^{13}$&$60.0\%$&$0.7\%$\\
&Y&$2.339\times10^{-2}$&$5.998\times10^{-9}$&$-3.106\times10^{-17}$&$1.818\times10^{-2}$&$1.193\times10^{-8}$&$-8.927\times10^{-11}$&$2.728\times10^{13}$&$43.6\%$&$0.8\%$\\
$ED$&J&$2.603\times10^{-2}$&$1.334\times10^{-8}$&$-1.669\times10^{-16}$&$1.799\times10^{-2}$&$2.797\times10^{-8}$&$-2.660\times10^{-10}$&$7.920\times10^{12}$&$67.9\%$&$1.0\%$\\
&H&$2.754\times10^{-2}$&$2.128\times10^{-8}$&$-4.023\times10^{-16}$&$1.386\times10^{-2}$&$4.463\times10^{-8}$&$-4.617\times10^{-10}$&$2.814\times10^{13}$&$92.6\%$&$1.0\%$\\
&K&$2.986\times10^{-2}$&$3.690\times10^{-8}$&$-1.325\times10^{-15}$&$1.522\times10^{-2}$&$7.915\times10^{-8}$&$-9.672\times10^{-10}$&$5.005\times10^{12}$&$90.4\%$&$0.9\%$\\
\hline
&Z&$3.336\times10^{-2}$&$4.571\times10^{-9}$&$-2.053\times10^{-17}$&$2.947\times10^{-2}$&$9.286\times10^{-9}$&$-6.791\times10^{-11}$&$8.570\times10^{12}$&$85.0\%$&$0.9\%$\\
&Y&$2.269\times10^{-2}$&$5.760\times10^{-9}$&$-3.139\times10^{-17}$&$1.781\times10^{-2}$&$1.182\times10^{-8}$&$-9.105\times10^{-11}$&$1.913\times10^{13}$&$71.8\%$&$1.1\%$\\
$ED_{(1)}$&J&$2.702\times10^{-2}$&$1.255\times10^{-8}$&$-1.625\times10^{-16}$&$1.797\times10^{-2}$&$2.746\times10^{-8}$&$-2.673\times10^{-10}$&$9.325\times10^{12}$&$86.6\%$&$1.4\%$\\
&H&$2.836\times10^{-2}$&$1.972\times10^{-8}$&$-3.861\times10^{-16}$&$1.358\times10^{-2}$&$4.338\times10^{-8}$&$-4.606\times10^{-10}$&$3.078\times10^{13}$&$94.9\%$&$1.2\%$\\
&K&$3.235\times10^{-2}$&$3.338\times10^{-8}$&$-1.217\times10^{-15}$&$1.600\times10^{-2}$&$7.568\times10^{-8}$&$-9.447\times10^{-10}$&$5.720\times10^{12}$&$94.2\%$&$1.1\%$\\
\hline
&Z&$2.699\times10^{-2}$&$3.624\times10^{-9}$&$-1.572\times10^{-17}$&$2.388\times10^{-2}$&$7.222\times10^{-9}$&$-5.186\times10^{-11}$&$1.008\times10^{13}$&$85.9\%$&$0.6\%$\\
&Y&$1.836\times10^{-2}$&$4.583\times10^{-9}$&$-2.423\times10^{-17}$&$1.445\times10^{-2}$&$9.258\times10^{-9}$&$-7.021\times10^{-11}$&$1.734\times10^{13}$&$69.6\%$&$0.9\%$\\
$ED_{(2)}$&J&$2.188\times10^{-2}$&$9.945\times10^{-9}$&$-1.245\times10^{-16}$&$1.449\times10^{-2}$&$2.145\times10^{-8}$&$-2.056\times10^{-10}$&$8.007\times10^{12}$&$87.3\%$&$1.1\%$\\
&H&$2.351\times10^{-2}$&$1.553\times10^{-8}$&$-2.871\times10^{-16}$&$1.103\times10^{-2}$&$3.393\times10^{-8}$&$-3.533\times10^{-10}$&$2.670\times10^{13}$&$95.6\%$&$1.1\%$\\
&K&$2.698\times10^{-2}$&$2.620\times10^{-8}$&$-9.019\times10^{-16}$&$1.268\times10^{-2}$&$5.969\times10^{-8}$&$-7.336\times10^{-10}$&$5.140\times10^{12}$&$94.5\%$&$1.3\%$\\
\hline

\end{tabular}
\end{table*} 

\begin{itemize}
 \item The dispersion even-parameters have a similar range as those found for 
 the standard deviation. Moreover, the majority of WVSC1 stars have values
 above the stochastic variations and therefore these parameters can be used in the same
 fashion as the standard deviation to discriminate variable stars from noise. 
 As expected, the diagram of $ED$ is equal to $ED_{M}$ and is being similar
 to $ED_{\mu}$.
 
\item The Strateva and the modified-Strateva functions show similar values for
almost all ranges of magnitude. The difference is a slope at lower magnitudes
(bright stars) found for the modified-Strateva function. This allows us to
reduce the mis-selection but we also remove some bright variable stars
that have small amplitude variations. We caution that Strateva and the
modified-Strateva functions can present an incorrect model for very faint
magnitudes since a small decrease in the dispersion is found. In these cases a 
magnitude limit can be adopted \citep[][]{Cross-2009}. 

\item The shape even-parameters give a good discrimination for many variable 
stars particularly for bright stars (see Fig.~\ref{fg_magdispshape}).
However, almost all faint stars (magnitudes greater than $\sim 16$) have values
near those found for stochastic variations. In this sense, the 
dispersion parameters are better than shape parameters at discriminating
non-stochastic variations since we can see a clearer separation among them for 
all ranges of magnitude. The shape parameters may be useful to discriminate
different kinds of light curve signatures and this will be addressed in a
future paper in this series.
\end{itemize}

In summary, the even-statistical parameters can be used in the same fashion as
previous parameters. The main goal of this paper is to study the criteria of selection
of variable stars from noise and meanwhile these parameters may be useful for
many other purposes in different branches of science and technology.

 \begin{figure}[htb]
  \centering
  \includegraphics[width=0.45\textwidth,height=0.40\textwidth]{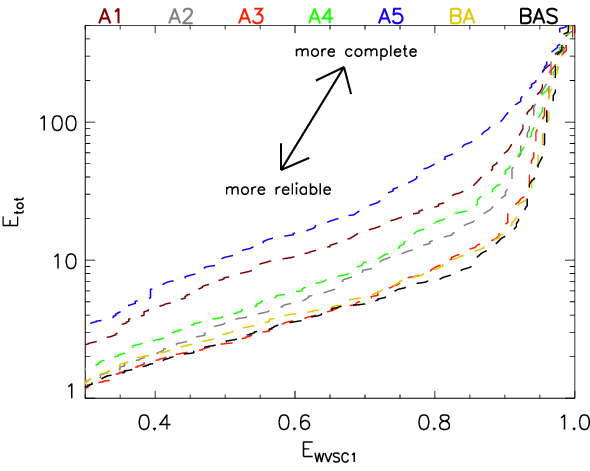}  
  \includegraphics[width=0.45\textwidth,height=0.40\textwidth]{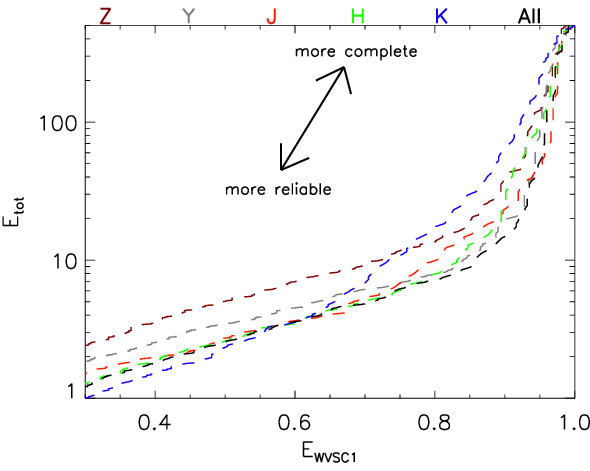}  

  \caption{Top panel shows $E_{tot}$ as a function of $E_{WVSC1}$ for all apertures and
  for the bottom panel for all individual wavebands and the combination
  (ZYJHK wavebands) using $BA$. Here the result for each photometric aperture
  and waveband are shown with different colours. $E_{WVSC1}$ decreases with
  $E_{tot}$ leading to a more reliable selection (fewer misclassifications) and
  vice versa.}
 \label{fg_etot}
\end{figure}

\subsection{Finding the best noise model}\label{sec_test_noisemodel}

We tested $82159$ models to find the best model
Strateva-modified function ($\zeta_{P} ( m )$ - see Eq. \ref{eq_stratevamod}).
All combinations in a range of three power terms varying from $4$ to $-4$ were
performed using a bin of $0.1$. This range covered all previous values used in
the noise model. The procedure was adopted to find the best model to fit the
standard deviation as a function of magnitude is similar to that used by
\citet[][]{Cross-2009}. The $EA_{M}$ and $ED_{M}$ are computed for bins with 
a width of $0.1$ magnitude or at least $100$ objects. For this step, we
only consider those stars with more than $20$ measurements. Next, we compute 
$\zeta_{P'} ( m )$ from a non-linear least-squares minimization using the 
Levenberg-Marquardt method \citet{Levenberg-1944,Marquardt-1963}. The WVSC1 catalogue
of variables represent $0.01\%$ of WFCAMCAL stars. However, they were removed from the sample before fitting
to get a better noise model,.

About $39\%$ of the models tested converge for all
statistical parameters and wavebands observed. The model with the
lowest $\chi^2$ in $ZYJHK$ wavebands was taken as the best noise model
given by Eq. \ref{eq_stratevamod}. Table \ref{tab_noisemodel} shows the
parameters obtained for both Strateva and Strateva-modified functions and the 
metric to measure the improvement or deterioration provided by the latter. The
dispersion of residuals ($G(R)$) is about $1\%$ lower than that found for
Strateva functions for almost all statistical parameters except for $\rm IQR$
and $\rm EIQR$. A similar behaviour is found for $G(\chi^2)$ where an improvement
of about $85\%$ is found. Indeed, the largest improvement is found for the K
filter.
However, a deterioration is found for $\rm IQR$ and $\rm EIQR$ in the $ZYJ$
wavebands.
Moreover, Strateva-modified functions do not turn down at faint magnitudes as the Strateva
function does sometimes (see Fig. \ref{fg_magdispshape}). The new noise model
provides a more restrictive cut for both the brightest and faintest stars.

\subsection{Testing photometric apertures and wavebands}\label{sec_testaper}

In order to test the dependence of the photometric aperture and extreme
measurements on the selection criteria the WFCAM, seven different analyses were performed:

\begin{itemize}
 \item \textbf{A1-5} Photometric measurements using a standard photometric
 aperture from 1 to 5 ($0.5\arcsec$, $\sqrt(0.5)\arcsec$, $1\arcsec$,
 $\sqrt(2)\arcsec$, and $2\arcsec$ radius, respectively)
 \item \textbf{BA} Photometric measurements using the best aperture 
 \citep[see][]{Cross-2009}
 \item \textbf{BAS} all measurements enclosed in $2\times
 ED_{\sigma\mu}$ about $EA_{M}$ of BA photometry are used. 
\end{itemize}

\begin{table*}[htbp]
 \scriptsize 
\caption[]{Efficiency metric $E_{tot}$, i.e. the ratio of the number of selected 
sources to the total number of WVSC1 variable stars, and $E_{WVSC1}$, i.e. the ratio
of the number of WVSC1 stars selected to the total number of WVSC1 stars, and
$\alpha$ values computed from $X_{f}(ED)$ for each waveband and for using all
$ZYJHK$ wavebands using $\beta = 4$. This efficiency metric was performed for different 
photometric apertures (A1-5), BA, and BAS.}\label{tab_ef}
 \centering 
 \begin{tabular}{| c | c | c c  | c c  | c c | c c | c c |  c c |}        
 \hline                
 \multicolumn{1}{|c}{}  &  \multicolumn{1}{|c}{}  & \multicolumn{2}{|c|}{Z } & \multicolumn{2}{|c|}{Y } & \multicolumn{2}{|c|}{J } & \multicolumn{2}{|c|}{H } & \multicolumn{2}{|c|}{K }  & \multicolumn{2}{|c|}{YZJHK } \\ 
 \hline
 & $E_{WVSC1}$ &  $\alpha$ & $E_{tot}$ &  $\alpha$ & $E_{tot}$& $\alpha$ & $E_{tot}$& $\alpha$ & $E_{tot}$& $\alpha$ & $E_{tot}$& $\alpha$ & $E_{tot}$ \\    
  \hline
  & 0.60 & 1.69 & 11.68 & 1.77 & 11.32 & 1.55 & 12.51 & 1.51 & 15.44 & 1.35 & 17.12 & 1.60 & 10.44 \\
  & 0.65 & 1.61 & 13.60 & 1.63 & 14.31 & 1.46 & 15.18 & 1.39 & 19.72 & 1.28 & 21.01 & 1.52 & 12.28 \\
  & 0.70 & 1.45 & 19.06 & 1.55 & 16.63 & 1.36 & 19.33 & 1.32 & 23.20 & 1.18 & 30.81 & 1.40 & 16.03 \\
A1  & 0.75 & 1.33 & 26.04 & 1.44 & 20.54 & 1.28 & 24.12 & 1.24 & 28.95 & 1.10 & 44.75 & 1.31 & 20.20 \\
  & 0.80 & 1.24 & 33.62 & 1.31 & 27.84 & 1.22 & 29.61 & 1.14 & 41.63 & 1.02 & 69.56 & 1.25 & 23.97 \\
  & 0.85 & 1.16 & 43.36 & 1.23 & 35.08 & 1.14 & 40.33 & 1.06 & 60.19 & 0.93 & 118.51 & 1.15 & 33.10 \\
  & 0.90 & 1.11 & 52.43 & 1.07 & 65.18 & 1.03 & 68.19 & 0.94 & 119.14 & 0.83 & 200.26 & 1.02 & 58.80 \\
 \hline
  & 0.60 & 1.80 & 9.86 & 1.89 & 5.94 & 1.67 & 6.43 & 1.68 & 6.68 & 1.50 & 8.08 & 1.72 & 4.86 \\
  & 0.65 & 1.65 & 12.61 & 1.76 & 7.43 & 1.54 & 8.89 & 1.54 & 9.28 & 1.39 & 11.77 & 1.64 & 5.87 \\
  & 0.70 & 1.48 & 17.53 & 1.65 & 9.28 & 1.48 & 10.48 & 1.44 & 12.01 & 1.29 & 17.18 & 1.49 & 8.43 \\
A2  & 0.75 & 1.40 & 20.98 & 1.51 & 12.78 & 1.39 & 13.98 & 1.38 & 14.33 & 1.21 & 25.03 & 1.40 & 10.90 \\
  & 0.80 & 1.31 & 26.42 & 1.42 & 16.34 & 1.30 & 19.28 & 1.26 & 21.74 & 1.11 & 42.00 & 1.33 & 13.63 \\
  & 0.85 & 1.22 & 35.34 & 1.30 & 23.81 & 1.25 & 23.49 & 1.17 & 32.43 & 1.04 & 64.04 & 1.25 & 18.19 \\
  & 0.90 & 1.08 & 60.37 & 1.14 & 45.97 & 1.12 & 43.38 & 1.07 & 55.76 & 0.95 & 110.79 & 1.15 & 28.06 \\
 \hline
  & 0.60 & 1.96 & 10.39 & 2.19 & 4.03 & 1.89 & 4.06 & 1.85 & 3.43 & 1.61 & 5.27 & 1.89 & 3.61 \\
  & 0.65 & 1.83 & 12.35 & 2.00 & 4.92 & 1.78 & 4.87 & 1.77 & 3.91 & 1.49 & 7.85 & 1.78 & 4.24 \\
  & 0.70 & 1.71 & 14.67 & 1.88 & 5.80 & 1.70 & 5.84 & 1.67 & 4.84 & 1.41 & 10.61 & 1.68 & 5.11 \\
A3  & 0.75 & 1.60 & 17.46 & 1.72 & 7.47 & 1.55 & 8.49 & 1.55 & 6.59 & 1.31 & 16.56 & 1.54 & 6.95 \\
  & 0.80 & 1.49 & 21.29 & 1.61 & 9.24 & 1.45 & 11.42 & 1.40 & 10.59 & 1.20 & 28.56 & 1.45 & 8.79 \\
  & 0.85 & 1.31 & 31.60 & 1.48 & 12.60 & 1.33 & 17.39 & 1.29 & 16.86 & 1.12 & 44.18 & 1.35 & 11.94 \\
  & 0.90 & 1.17 & 47.78 & 1.34 & 19.45 & 1.26 & 23.33 & 1.18 & 29.47 & 0.99 & 95.40 & 1.25 & 17.89 \\
 \hline
  & 0.60 & 1.89 & 14.00 & 2.14 & 6.61 & 1.83 & 7.56 & 1.79 & 6.21 & 1.49 & 12.50 & 1.85 & 5.81 \\
  & 0.65 & 1.75 & 16.91 & 2.03 & 7.61 & 1.72 & 9.43 & 1.67 & 7.93 & 1.39 & 17.76 & 1.73 & 7.19 \\
  & 0.70 & 1.61 & 21.36 & 1.82 & 10.21 & 1.64 & 11.24 & 1.58 & 9.73 & 1.30 & 25.31 & 1.59 & 9.67 \\
A4  & 0.75 & 1.51 & 25.59 & 1.62 & 14.64 & 1.48 & 16.70 & 1.46 & 13.61 & 1.23 & 33.68 & 1.48 & 12.63 \\
  & 0.80 & 1.36 & 34.97 & 1.53 & 18.06 & 1.38 & 22.76 & 1.36 & 19.03 & 1.16 & 46.39 & 1.36 & 18.34 \\
  & 0.85 & 1.20 & 53.13 & 1.41 & 24.63 & 1.27 & 33.60 & 1.27 & 27.13 & 1.07 & 71.66 & 1.29 & 23.48 \\
  & 0.90 & 1.08 & 77.22 & 1.26 & 40.16 & 1.15 & 55.79 & 1.09 & 66.75 & 0.92 & 148.32 & 1.16 & 40.53 \\
 \hline
  & 0.60 & 1.70 & 24.95 & 1.90 & 15.88 & 1.65 & 19.50 & 1.55 & 18.80 & 1.30 & 37.04 & 1.64 & 15.61 \\
  & 0.65 & 1.60 & 29.20 & 1.77 & 19.38 & 1.54 & 24.54 & 1.43 & 25.66 & 1.23 & 47.27 & 1.55 & 18.93 \\
  & 0.70 & 1.44 & 39.15 & 1.62 & 25.08 & 1.45 & 30.33 & 1.33 & 34.90 & 1.17 & 59.55 & 1.46 & 23.60 \\
A5  & 0.75 & 1.32 & 50.30 & 1.43 & 37.38 & 1.34 & 40.88 & 1.24 & 48.46 & 1.11 & 76.34 & 1.32 & 34.71 \\
  & 0.80 & 1.20 & 67.13 & 1.31 & 50.83 & 1.22 & 60.07 & 1.20 & 56.69 & 1.01 & 116.02 & 1.22 & 48.46 \\
  & 0.85 & 1.07 & 96.87 & 1.21 & 69.58 & 1.12 & 86.07 & 1.10 & 86.82 & 0.95 & 147.82 & 1.13 & 68.59 \\
  & 0.90 & 0.98 & 130.19 & 1.12 & 94.67 & 1.03 & 121.97 & 0.97 & 153.50 & 0.83 & 229.28 & 1.03 & 106.21 \\
 \hline
  & 0.60 & 2.00 & 10.90 & 2.23 & 5.30 & 1.98 & 4.30 & 1.95 & 4.54 & 1.68 & 4.49 & 1.96 & 4.06 \\
  & 0.65 & 1.85 & 13.13 & 2.11 & 5.91 & 1.83 & 5.32 & 1.83 & 5.34 & 1.56 & 6.12 & 1.88 & 4.45 \\
  & 0.70 & 1.76 & 14.80 & 1.96 & 6.85 & 1.75 & 6.07 & 1.74 & 6.08 & 1.44 & 8.89 & 1.74 & 5.47 \\
AB  & 0.75 & 1.67 & 16.76 & 1.76 & 8.78 & 1.59 & 8.33 & 1.64 & 7.13 & 1.33 & 13.30 & 1.59 & 7.03 \\
  & 0.80 & 1.50 & 21.88 & 1.70 & 9.58 & 1.45 & 11.98 & 1.47 & 10.52 & 1.24 & 19.99 & 1.50 & 8.49 \\
  & 0.85 & 1.33 & 30.98 & 1.54 & 12.51 & 1.37 & 15.45 & 1.34 & 15.56 & 1.14 & 33.75 & 1.40 & 10.93 \\
  & 0.90 & 1.17 & 47.82 & 1.35 & 20.45 & 1.24 & 25.84 & 1.21 & 26.20 & 1.01 & 74.20 & 1.29 & 15.47 \\
 \hline
  & 0.60 & 2.09 & 6.98 & 2.29 & 4.51 & 2.02 & 3.61 & 1.98 & 3.58 & 1.68 & 3.49 & 2.02 & 3.57 \\
  & 0.65 & 1.95 & 7.97 & 2.08 & 5.31 & 1.93 & 3.98 & 1.81 & 4.35 & 1.55 & 4.77 & 1.85 & 4.35 \\
  & 0.70 & 1.82 & 9.42 & 1.94 & 6.08 & 1.71 & 5.39 & 1.72 & 4.91 & 1.45 & 6.75 & 1.76 & 4.91 \\
ABS  & 0.75 & 1.66 & 11.71 & 1.83 & 6.82 & 1.60 & 6.73 & 1.60 & 5.95 & 1.34 & 10.50 & 1.60 & 6.19 \\
  & 0.80 & 1.57 & 13.42 & 1.72 & 7.89 & 1.46 & 9.84 & 1.48 & 7.86 & 1.25 & 16.56 & 1.53 & 7.05 \\
  & 0.85 & 1.39 & 19.32 & 1.53 & 10.71 & 1.34 & 14.71 & 1.33 & 12.80 & 1.15 & 29.54 & 1.43 & 8.71 \\
  & 0.90 & 1.16 & 40.05 & 1.32 & 19.24 & 1.24 & 22.95 & 1.20 & 23.74 & 1.03 & 63.71 & 1.27 & 14.36 \\
 \hline

\end{tabular}
\end{table*} 

In these analysis, the measurements with flags greater than 256 were removed. 
The third aperture (A3), corresponds to the default $1\arcsec$ aperture, where the
radius is slightly larger than the typical seeing FWHM, so an aperture centred
on a point-source should contain $>95\%$ of the light in the ideal Gaussian 
case; in reality there is much more light in lower surface brightness wings. Increasing the
aperture size increases the amount of signal, but at the expense of
increasing the amount of sky too, such that the signal-to-noise
decreases. Decreasing the aperture reduces the signal too much, also reducing
the signal-to-noise ratio. Usually A3 gives the optimal signal-to-noise, but
sometimes, nearby stars can affect the measurements by adding an additional
noise component from deblending images that relies on some imperfect modelling,
and selecting a smaller aperture that includes less signal from the
neighbor gives better results, which is why a variable aperture was selected by
\citet{Cross-2009}.

Figure \ref{fg_etot} shows the result for different 
apertures (top panel) and different wavebands (bottom panel). The BAS returns the 
best results, i.e. the lowest values of  $E_{tot}$ for all values of 
$E_{WVSC1}$. The BAS approach allows us to achieve a better discrimination
of variable stars from noise  (see Table \ref{tab_ef}). 
 It is mainly noted for those dispersion parameters that take into
account the square in their definition, such as $ED_{\sigma_\mu}$ and $ED_{\sigma_m}$.
On the other hand, the BAS approach can also lead to mis-selections of binary stars that have few measurements at
the eclipse, for instance (see Sect.~\ref{sec_testaper} for more details). The 
number of stochastic variations decreases a lot but it also means that we could 
miss some variable stars. On the other hand, the efficiency levels for
different wavebands vary significantly (see Table \ref{tab_ef}).  
The best result was found for the $J$ waveband rather that for the $Z$ and $K$ 
wavebands. The efficiency decrease found for the $K$ waveband is related to
the decrease of signal-to-noise, while for the $Z$ waveband we find that the $c_0$
in Eq. \ref{eq_stratevamod} is significantly higher ($\sim0.023$; cf. 
$\sim0.014$ for $Y$, $J$, $H$, $K$), which suggests greater variations in the photometry across each detector, since simple offsets in the zero point would be corrected by the
recalibration carried out by \citet{Cross-2009}. Calibrating the $Z$ and $Y$ bands was
trickier than the J, H, K bands because the calibration is extrapolated from 2MASS $J$, 
$H$, $Ks$ \citep[see][]{Hodgkin-2009} and more susceptible to extinction, 
particularly in the $Z$ band, which can vary on small scales in star forming 
regions. Indeed, $32$ WVSC1 stars were found in highly reddened
regions ($Z - K > 3$) indicating that such effects can be present in WFCAM
data.

\begin{figure*}[htb]
  \centering
  \includegraphics[width=0.9\textwidth,height=1.\textwidth]{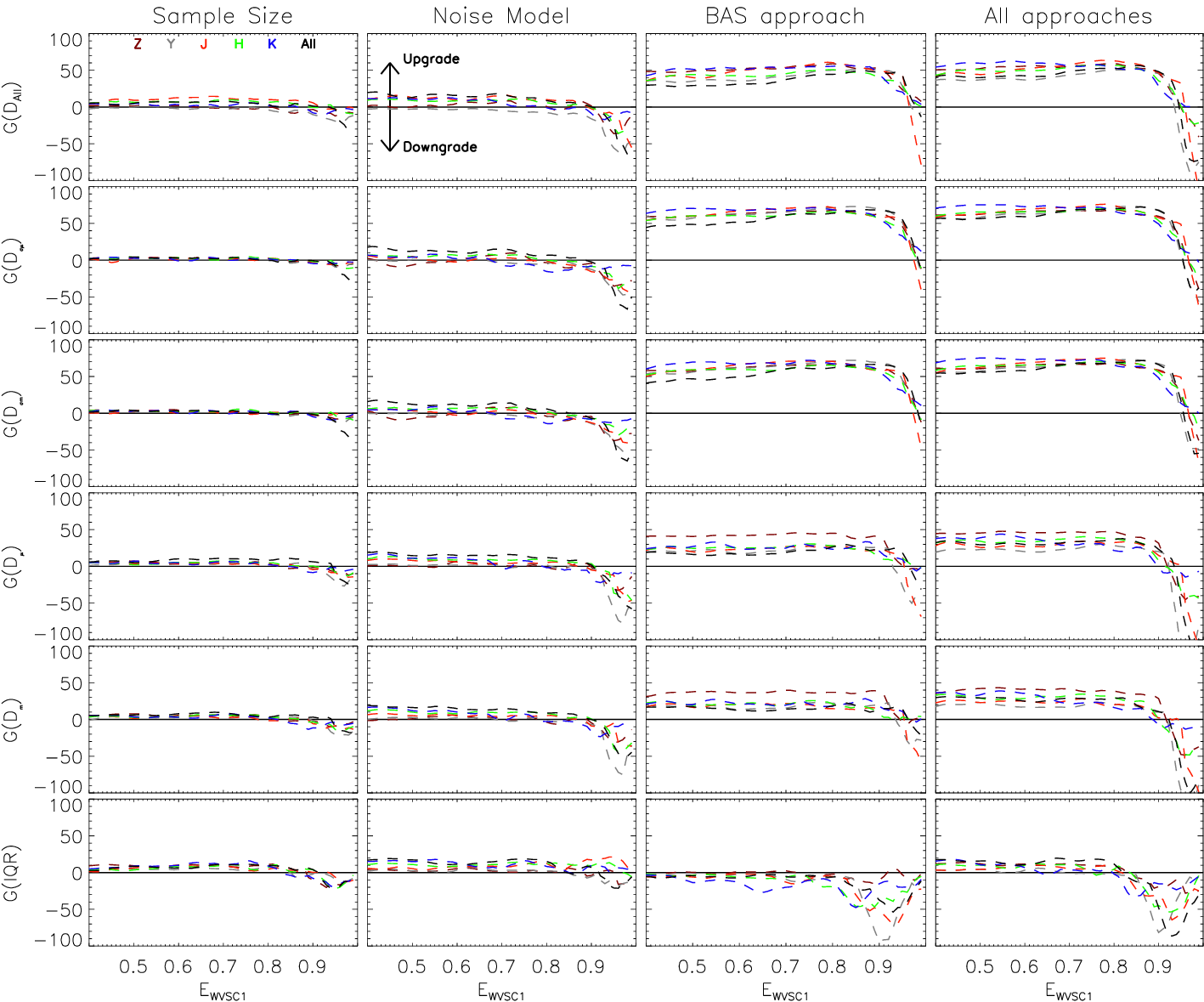}  

  \caption{$G$ vs. $E_{WVSC1}$ using different
  approaches  (see \ref{sec_testupgrade}). The approach used is named above in
  each of the upper diagrams. The colours indicate the results for different
  filters ZYJHK (brown, grey, red, green, and blue lines respectively) and the combination 
  of results found in all bands (black lines). The same colours were also used in
  Fig. \ref{fg_etot} (bottom panel). }
 \label{fg_testimprovements} 
\end{figure*}

\subsection{Analysis of improvements}\label{sec_testupgrade}

Section \ref{sec_test_noisemodel} discusses the
improvements made using the Strateva-modified function.
These improvements have smaller $\chi^2$ than the original Strateva function, which indicates
a better noise model estimation. However, this does not inform us about the 
improvements to the selection of variables. Therefore,  to measure  the
improvements or deteriorations provided by each step of our analysis the metric $G$ was computed for $E_{tot}$ and $E_{WVSC1}$ using four different approaches as follows;

\begin{itemize}
 \item \textbf{Even-statistic - } The results are computed
 from standard dispersion parameters in comparison with their respective
 counterpart even-dispersion parameter for BA photometry (see Sect.
 \ref{sec_testaper}).
 \item \textbf{Sample size - } The results with and 
 without sample size corrections for BA photometry.
 \item \textbf{Noise model - } The results using the
 Strateva versus Strateva modified functions for BA photometry.
  \item \textbf{BAS approach - } The results for BA
  photometry versus with those computed for the BAS approach.
 \item \textbf{All - }  The results computed from the 
 previous dispersion parameter using the Strateva function without sample size 
 corrections for BA photometry versus their respective even 
 dispersion parameter using sample size correction, Strateva modified functions,
 and the BAS approach.
\end{itemize}

\begin{figure*}[htb]
  \centering
  \includegraphics[width=0.49\textwidth,height=0.45\textwidth]{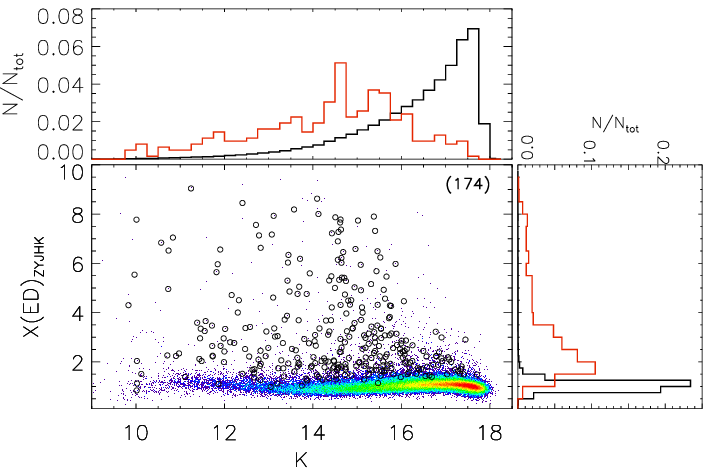}  
  \includegraphics[width=0.49\textwidth,height=0.45\textwidth]{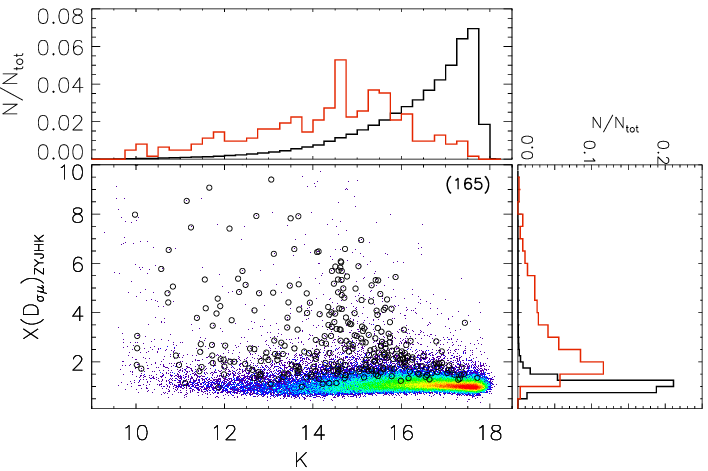}  

  \caption{$X$ indices using all wavebands $X_{ZYJHK}$ for
  $ED$ and $D_{\sigma \mu}$. The $X(ED)$ index was computed using the 
  Strateva function without sample size correction for BA photometry, while 
  $X(ED)$ was computed using the Strateva modified function with sample size
  correction for the BAS approach. The histogram of the entire sample (black lines)
  and WVSC1 stars (red lines) are shown at the top right. The WVSC1 
  stars are also represented by open black circles and the maximum number of
  sources per pixel is shown in brackets. }
 \label{fg_index} 
\end{figure*}

This analysis allows us verify if and how much each approach upgrades 
or downgrades the selection criteria compared with previous ones. Moreover, the
following combination of dispersion parameters were also analyzed:

\begin{equation}
  D_{All} = \left[ D_{\sigma\mu},D_{\sigma m},D_{\mu},D_{m} \right]
 \label{eq_dall}        
\end{equation}

and

\begin{equation}
  ED_{All} = \left[ED_{\sigma\mu},ED_{\sigma m},ED_{\mu},ED_{m} \right]
 \label{eq_edall}        
\end{equation}

These combinations provide the same number of dispersion
parameters using either the standard statistics or the even-statistics and hence
give a better comparison of their efficiency level. Where the weight
$\omega_{P_{j}}$ was adopted as the inverse of $ED_{\sigma\mu}$ for each
$D_{j}$. The same tests were performed using single dispersion parameters 
to verify if their combination provides better results.

Figure \ref{fg_testimprovements} shows a comparison between
previous and current approaches. The even-statistical parameters are on average
better than the standard statistical parameters (see Sect. \ref{sec_Accurate}). The results for
real data show a fluctuation of about $1\%$ on $G(E_{tot})$ values. This is expected
since the improvements to the estimation of standard statistical parameters 
only occur for those distributions that have odd numbers of measurements and 
decreases quickly with the number of measurements. Moreover, we also observed 
that the improvements vary from the $Z$ to $K$ waveband since the infrared light
curves usually have smaller amplitudes than optical wavebands and therefore the improvement is
more evident. The following behaviours are also observed;

\begin{itemize}
\item The $G(E_{tot})$ for the sample size correction
varies from $2\%$ to $7\%$ for $E_{WVSC1}$ less than $\sim 0.8$. 
The $E_{WVSC1}$ stars outside of this limit have an $X_{ED}$ (see Fig.
\ref{fg_index}) less than $1.5$. Indeed, the $X$ variability indices are
approximately a measurement of the signal-to-noise ratio so this indicates an
improvement in the signal-to-noise greater than $1.5$ for $E_{WVSC1}<0.8$ and a
deterioration of about $7\%$ otherwise.

\item The improvement provided by the Strateva modified
function can reach $G(E_{tot}) \simeq 22\%$. It only improves the selection for
$E_{WVSC1}$ lower than $\sim 0.9$ similar to that found for the sample size 
correction. Indeed, the Strateva modified function provides a fluctuation of 
about few percent of improvement or diminishment for Z and Y wavebands. The
increase to the total number selected provided by the sample size correction and
noise model means a reduction of misclassification but this also hinders the
detectability of variable stars of lower amplitudes that are mainly found at fainter
magnitudes.

\item The BAS approach provides the largest improvement to
the selection criteria for all dispersion parameters tested except IQR. The
definition of IQR takes into account $75\%$ of the distribution and hence the BAS
approach to IQR provides a second reduction on the data used. Therefore the BAS
approach to IQR is not appropriate. On the other hand, the BAS approach is
suitable for all dispersion parameters analyzed since the maximum improvement
found is about $73\%$, where $D_{\sigma \mu}$ and $D_{\sigma m}$ 
have the largest improvements.

\item The total improvement is dominated by the
improvement from the BAS approach since the maximum improvement is not so
different to that found for BAS approach. Indeed, the BAS approach leads to a
constant improvement until $E_{WVSC1}\simeq 0.95$. The decrease observed for
values higher than that worsens when the sample size correction and the Strateva
modified function are added.

\item We also perform the selection using the previous
standard procedure to select variable stars using non-correlated data, i.e.
select all sources with an magnitude RMS above n times sigma above the noise 
model function. We compute the standard deviation and the X index for the K
waveband using BA photometry. At one sigma above the Strateva function $\sim 81\%$ of
WVSC1 stars are selected but at the expense of an $E_{tot} \simeq 103$. This
$E_{tot}$ value is $\sim 5.2$ times larger than that found using our approach
for the K waveband and $\sim 12$ times that considering all wavebands (see Table
\ref{tab_ef}). This means that the modified-Strateva function joined with our
empirical approach (see Eqs.~\ref{eq_bcutoff1} and \ref{eq_stratevamod})
and statistical weights (see Sect.~\ref{sec_Sumary_tests}) increases the 
selection efficiency by about $\sim 520\%$.

\item The performance of previous and even-statistical 
parameters are very similar when we use the sample size correction and
Strateva-modified function with the BAS approach. Indeed, the efficiency level
for EIQR is optimized if only the Strateva-modified function and sample size 
correction for BA photometry is used.

\item The performance obtained from single statistical 
parameters in comparison with those found for $ED_{All}$ or $D_{All}$ are very
similar. Therefore the combination of several statistical parameters does not
provide an improvement according to our results. Moreover, the combination with more
statistical parameters was performed but no improvement was found.

\item The largest improvement is found when all wavebands
are combined. This returns a set of potential variable stars of about $2.1$, for
$E_{WVSC1}\simeq 0.8$ and $4.9$, for $E_{WVSC1}\simeq 0.9$, times smaller  than
that found for single wavebands.

\end{itemize}

The approaches proposed in the current work provide
reasonable improvements to the selection criteria. All steps of our approach
were tested allowing us to identify which parameters are improved and the
range of $E_{WVSC1}$ over which these improvements are valid. Such analyses
allow us to define the best way to select variable stars with statistical parameters.

\begin{figure*}[htb]
  \centering
  \includegraphics[width=0.49\textwidth,height=0.45\textwidth]{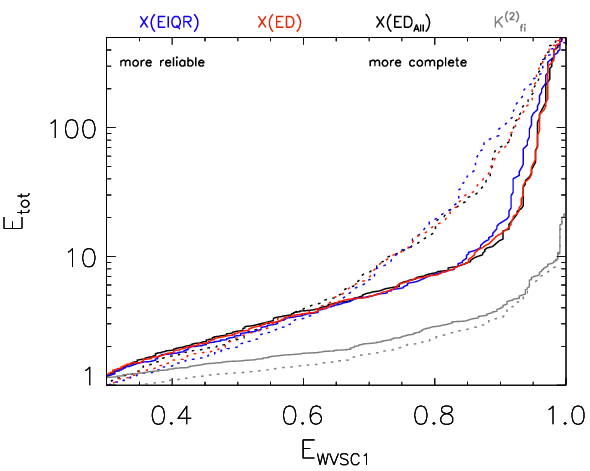}  
  \includegraphics[width=0.49\textwidth,height=0.45\textwidth]{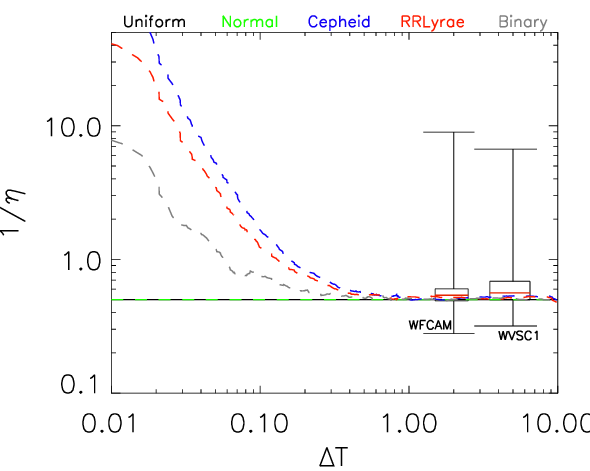}  

  \caption{$E_{tot}$ vs. $E_{WVSC1}$ (left panel) for
  the best selection criteria and $1/\eta$ as a function of time interval mean
  among the measurements $\Delta T$ (right panel). In the left panel, the
  statistical parameters $X(EIQR)$ (blue lines), $X(D_{\sigma\mu})$ (red lines),
  and $X(ED)$ (dark lines) for $ZYJHK$ (full lines) and $K$ (dashed lines)
  wavebands, respectively. In the same diagram the the results found
  for $K_{fi}^{(2)}$ using the mean (full grey line) and even-mean (dashed grey
  line), respectively, are also plotted. The results for normal, uniform, Ceph, RR, and EB
  simulated distributions are shown in the right panel.
  The colours indicate different parameters or distributions analyzed, which are indicated at the top of each diagram. 
  In the right panel we show the how $1/\eta$ varies with the selection of
  $\Delta T$. The box plot indicates the quartiles of WFCAM and WVSC1 samples,
  where each box encloses $50\%$ of the sample under the grouping algorithm that
  defines $\Delta T$ in \citet[][]{Sokolovsky-2017}.}
 \label{fg_neta} 
\end{figure*}

\subsection{Improvements on correlated indices}\label{sec_ncorcor}

The flux independent variability indices ($K_{fi}^{(s)}$) proposed by 
us \citep[for more details see][]{FerreiraLopes-2015wfcam,FerreiraLopes-2016I} are 
not dependent on the amplitude signal since they only use the correlation
signal. However, they are dependent on the mean value. Therefore, the
correlation values computed using even-averages are more accurate than those
computed using mean values since the even-mean gives a value closer to the true
centre (see Sect. 3.1). As a result, the $E_{tot}$ values
presented in Table~\ref{tab_efz} are reduced by about $\sim 18\%$ compared
to those found in Table~2 of paper I.  This reduction 
is related with to the sources that have few correlations. 
Such an improvement is almost constant for $E_{WVSC1} < 0.90$ (see Fig.
\ref{fg_neta}) and it is this high because a small variation in the number of
positive correlations provides a substantial improvement for the
$K_{fi}^{(s)}$ indices. For instance, a single correlation can create a variation of about
$20\%$ of $K_{fi}^{(2)}$ indices if there are only five correlation
measurements. Therefore, a better mean value provides a strong correction on
$K_{fi}^{(s)}$ indices with few correlations.

We also tested the correlated indices using BAS, i.e. all measurements
enclosed in $2\times ED_{\sigma\mu}$ about $EA_{m}$ of BA photometry
(see Sect. ~\ref{sec_testaper}). The results are not that different from
those found for $E_{WVSC1} < 0.85$ (see Table~\ref{tab_efz}), while for 
$E_{WVSC1} > 0.85$ we found an $E_{tot}$ about $40\%$ higher. The measurements 
related to eclipsing binary stars are removed when we use BAS. The correlated 
and non-correlated indices can fail for low signal-to-noise variations and for non-contact binaries with few 
measurements at the eclipses.

The WFCAMCAL database allows us to compute correlated indices that have a number
of correlations greater than $N_{2}^{(min)}$ for about $\sim 94\%$ of
data. Variable stars with fewer correlations or not previously detected will
be explored in the next paper of this series, in which we will propose a new
periodicity search method and study selection criteria to produce a
cleaner sample.

\begin{table}[htbp]
\caption[]{Efficiency metric $E_{tot}$, $E_{WVSC1}$, and $\alpha_{cor}$
values computed from the analysis of BA  photometry for $K_{(fi)}^{(2)}$ and
$K_{(fi)}^{(3)}$. $\alpha_{cor}$ are values related to Eq.~16 of
\citet[][]{FerreiraLopes-2016I}. }\label{tab_efz}
 \centering 
 \begin{tabular}{| c |  c c  | c c  |}        
 \hline                
 \multicolumn{1}{|c}{}  &  \multicolumn{2}{|c}{s = 2}  & \multicolumn{2}{|c|}{s = 3} \\ 
 \hline
 $E_{WVSC1}$ &  $\alpha_{cor}$ & $E_{tot}$ &  $\alpha_{cor}$ & $E_{tot}$ \\    
 0.60 & 0.21 & 1.44 & 0.30 & 0.86 \\
 0.65 & 0.22 & 1.54 & 0.33 & 0.97 \\
 0.70 & 0.24 & 1.77 & 0.35 & 1.11 \\
 0.75 & 0.25 & 1.97 & 0.38 & 1.31 \\
 0.80 & 0.27 & 2.27 & 0.40 & 1.55 \\
 0.85 & 0.28 & 2.74 & 0.43 & 1.92 \\
 0.90 & 0.30 & 3.38 & 0.46 & 2.65 \\

\hline 
\end{tabular}
\end{table}

\section{Summary of recommendations}\label{sec_thebestway}

Reliable selections become more important than complete 
selections of variable stars when confronted by a very large amount of
photometric data. Visual inspection is usually performed to designate if an
object is a variable star or not \citep[e.g.][]{Pojmanski-2005,Graczyk-2011,
DeMedeiros-2013,FerreiraLopes-2015wfcam,FerreiraLopes-2015cycles,
FerreiraLopes2015mgiant,Song-2016}. This is accomplished even if good filtering 
is performed to remove image artefacts, cosmic ray hits, and point spread function 
(PSF) wings of a bright nearby objects \citep[e.g.][]{Fruchter-2002,Bernard-2010,
Denisenko-2011,Ramsay-2014,Desai-2016}. This is because stochastic and
non-stochastic variations do not look that different from the viewpoint of
statistical and correlated indices, especially for low signal-to-noise data.
Indeed, at the end of this project we aim to propose a non-supervised procedure that
allows us get an unbiased sample from analyzing a large data set, i.e. without
performing visual inspection.

Recently new statistical parameters were proposed
that include the error bars. These may improve the statistical parameters if the
error bars are well estimated. However, they can also increase the
uncertainties because it is common to find outliers with smaller error bars.
The performance of many of these parameters were recently tested by
\citet[][]{Sokolovsky-2017}. The authors used as test data, a sample with $127
539$ objects and more than $40$ epochs of which $1251$
variable stars were confirmed among them. The limit in the number of
measurements gives a straightforward comparison with surveys such as PanSTARRS
(with about 12 measurements) and the VVVX that will observe fewer epochs than VVV, but is still in 
the range between $25$ to $40$. The authors set the $1/\eta$ index as the best
way to select variable stars, but this is not true if the epoch
interval ($\Delta T$) is large.
Figure \ref{fg_neta} (right panel) shows the variation of the $1/\eta$ index as a function of
$\Delta T$. As you can verify, the separation between stochastic (uniform and
normal distributions) and variable stars become more evident only
for $\Delta T < 0.1$. The grouping of observations as defined for the $1/\eta$
index is in a single band, and hence $\Delta T \simeq 17.5$d for WVSC1 stars for
single wavebands. Therefore $1/\eta$ index is not suitable for select variable
stars from noise in the WFCAM database.
The WFCAM database was analyzed by correlated indices because the multi-band
observations provide a large number of measurements taken in intervals of 
$\Delta T \simeq 0.01$d. Unlike statistical parameters the correlated indices
can be computed using multi-wavebands and this is the best way to calculate
these indices in this case. Indeed, more than $50\%$ of WVSC1 variable stars could be
missed if the $1/\eta$ was adopted to analyze the WFCAMCAL database. Such
results are in agreement with those found by \citealt[][see Fig. 2 Sect. 
4.1]{FerreiraLopes-2016I}, where the authors performed this analysis using
$K_{fi}^{(s)}$ indices. Indeed, $\Delta T < 0.1$ was found for the current test 
data because the variable stars simulated (Cepheid, RRlyrae, and eclipsing
binary) have a variability period equal to 1. The parameter $\Delta T$ is not set to choose
which variability indices must be used to performed variability analysis.
The analyses of correlated observations \citep[][see Sect.
4.3]{FerreiraLopes-2016I} is mandatory to determine whether correlated
indices can be used and to set $\Delta T$. \citet[][]{Sokolovsky-2017} did not
take into account  our correlated indices \citep[][]{FerreiraLopes-2016I},
which have a well-defined limit and a high accuracy for only a few correlated
measurements. These indices only combined those measurements that provide good 
information about statistical correlation. Indeed, many variable stars could be missed.
The confidence correlated indices only can be computed if $\Delta T$ is a small
fraction of the variability period, such results are in agreement with those
results found by \citep[][]{FerreiraLopes-2016I}. This aspect limits a 
straightforward comparison between correlated and non-correlated indices 
performed by the authors.

Figure \ref{fg_neta} (left panel) shows a summary of our
efforts to provide the best way to select variables in correlated and 
non-correlated data. The values $K_{fi}^{(s)}$ correlated indices are more efficient than
previous correlated indices and should be adopted in the case in which correlated
indices can be calculated sensibly, but aspects of this still need to be
tested, especially in systems where the correlation order and number of 
permutations are very low. The flux independent indices are 
weakly dependent on magnitude but are strongly dependent on the time interval
among correlated measurements, such indices should be used when the observations have a 
natural correlation interval that is shorter than the typical epoch interval
(for more details, see Sec 4.3 Paper I). Indeed, a large number of variable
stars can be missed if this is not taken into account.

The discrimination of variable stars from noise is better 
distinguished using correlated indices than non-correlated indices and so these
should be adopted when they are available (see Sect.~ \ref{fg_neta}) otherwise
$X_{f}$ indices can be used (see Fig.~\ref{fg_neta}). Indeed, this also
determines how we may best perform photometric observations to maximize the
performance of selection criteria. A combination of these observations can be used but it is 
not mean a high performance. The better selection performed by correlated
indices is well known and therefore the correlated indices may be adopted to 
achieve a smaller mis-selection rate. Using all the above, the following set of procedures is
recommended as the best way to select variable stars:

\begin{table*}[htbp]
 \scriptsize 
\caption[]{$E_{\rm WVSC1}$ and $E_{\rm tot}$ values for all dispersion parameters analyzed using $\rm ZYJHK$ wavebands. The BAS approach is used for all parameters other than for $\rm IQR$ and $\rm EIQR$.}\label{tab_etotall}
 \centering 
 \begin{tabular}{| c | c |  c |  c | c | c | c | c | c | c | c | c | c | c | c | c |}        
 \hline                
 \multicolumn{1}{|c}{$E_{\rm WVSC1}$}  &  \multicolumn{15}{|c|}{$E_{tot}$} \\ 
 \hline               
  & $ED_{\sigma_\mu}$& $D_{\sigma_\mu}$& $ED_{\sigma_m}$& $D_{\sigma_m}$& $ED_{\mu}$& $D_{\mu}$& $ED_{m}$& $D_{m}$& $\rm EIQR$& $\rm IQR$& $ED$& $ED_{(1)}$& $ED_{(2)}$&$D_{\rm All}$&$ED_{\rm All}$\\
  \hline 
0.60& 3.80& 3.80& 3.91& 3.94& 3.61& 3.61& 3.49& 3.49& 3.90& 3.77& 3.57& 3.72& 3.73& 3.75& 3.75\\
0.65& 4.44& 4.44& 4.53& 4.57& 4.25& 4.25& 4.30& 4.29& 4.79& 4.72& 4.35& 4.40& 4.42& 4.32& 4.32\\
0.70& 5.21& 5.21& 5.25& 5.31& 4.98& 4.98& 4.78& 4.77& 5.38& 5.25& 4.91& 5.28& 5.29& 5.17& 5.19\\
0.75& 6.36& 6.36& 6.49& 6.40& 6.24& 6.24& 6.07& 6.06& 6.42& 6.23& 6.19& 6.18& 6.07& 6.23& 6.23\\
0.80& 7.72& 7.72& 7.72& 7.90& 7.17& 7.15& 6.92& 7.03& 7.33& 7.45& 7.05& 7.61& 7.58& 7.30& 7.29\\
0.85& 9.50& 9.50& 9.83& 9.88& 8.93& 9.11& 8.60& 8.57& 9.36& 9.37& 8.71& 9.35& 9.02& 8.89& 8.88\\
0.90& 13.70& 13.70& 13.58& 13.63& 14.23& 14.19& 15.27& 15.25& 14.82& 15.09& 14.36& 13.68& 13.15& 13.10& 13.10\\
  \hline 

\end{tabular}
\end{table*}

\begin{itemize}
  \item A histogram of the interval between observations 
 must be analyzed to define if correlated indices can be used 
 \citep[see Fig. 3 of][]{FerreiraLopes-2016I}. The approach used to perform the 
 variability analysis may only be chosen after examining the time interval among
 the measurements. The measurements used to compute $K_{fi}^{(s)}$ correlated
 indices must be correlated over a fraction of the minimum variability period.
 The parameter $K_{fi}^{(s)}$ has the highest performance among the correlated indices 
 analyzed and hence this should be adopted as the main tool to select variable stars
 using correlated data. Moreover, the even-mean should be used instead of the mean to compute $K_{fi}^{(s)}$ indices in order to improve the correlation estimation.
  \item A minimum of 5 correlated measurements must be
  adopted as the limit to discriminate variable stars from noise using
  correlated indices \citep[see Eq. 14 of Sect 4.1 of ][]{FerreiraLopes-2016I}. 
  \item A constant cut-off value may be adopted if you can 
  consider all time series on the same basis independent of the number of
  correlations. A cut-off using the number of correlations provides a better
  selection and therefore should be adopted if there are a reasonable number of
  correlations (more than 10). Indeed, we suggest that correlated indices
  are only calculated for stars that have a number of correlations greater than
  $10$.
  This increases the reliability of the correlated indices estimation and 
  allows those stars with few correlated measurements to be analyzed by statistical
  parameters. Moreover a higher order of correlated variability indices may be
  adopted if more than 2 measurements are available in each correlation
  interval.
 \item   The $X_{f}$ index may be used for time series 
 with less than $10$ correlated measurements. We must combine the information of
 all wavebands if multiwavelength data is available. This reduces the
 misclassification rate by about $680\%$ (see Sect. \ref{sec_testupgrade}). 
 A single dispersion parameter must be used to decrease the running time  since the performances for a combination of dispersion parameters is similar (see Table \ref{tab_etotall}).
 The $X_{f}(ED)$, $X_{f}(ED_{\mu})$, or $X_{f}(ED_{m})$ have performance in between $X_{f}(\rm EIQR)$ and $X_{f}(ED_{All})$  for $E_{WVSC1} < 0.85$ and otherwise better than $X_{f}(\rm EIQR)$ (see Fig. \ref{fg_neta}).
 Indeed, $ED$, $ED_{\mu}$, $ED_{m}$, or $\rm EIQR$ are not defined  using
 squares and so they are less affected by outliers. On the one hand,
 the sources nearby the noise model are better using those parameters defined with squares. The BAS approach must not be used if the $\rm EIQR$ parameter is used as the selection criteria.  
 The $X_{f}(ED)$, $X_{f}(ED_{\mu})$, $X_{f}(ED_{m})$, or their combination may be adopted
 to get a reliable sample ($E_{\rm WVSC1} \sim 0.85$) since they 
 have better performance on average than all dispersion parameters tested. On the other hand,
 $X_{f}(ED_{All})$ or $X_{f}(D_{All})$ may be adopted to get a complete sample once it has a  better performance for $E_{\rm WVSC1} \succeq 0.85$.
 \item A cut-off dependent on the number of measurements may
 be used as a parameter to select variable stars (see Eq. \ref{eq_bcutoff1}).
 \item The sample selected by correlated or non-correlated
 indices is not unbiased, i.e. several stochastic variations are enclosed 
 in this selection. The identification of periodic or aperiodic signals may be 
 performed by period finding methods. That will be addressed in forthcoming
 papers of this project.
\end{itemize}

\section{Conclusions}\label{sec_Conclusions}

Statistical parameters were analyzed as a tool to discriminate variable stars from noise. 
We observe that statistics based on an even number of measurements
provide better estimations of statistical parameters. Therefore, we propose 
even-statistics, where only even numbers of measurements are considered.
The even-averages gave better results than current averages for many of
distributions analyzed.
Therefore the previous shape and dispersion parameters were tested using even-averages. 
Next, seven unbound statistical parameters are proposed; i.e. they
are independent of the average. We propose 16 new statistical parameters are proposed in total. These parameters enlarge our inventory of tools to identify non-stochastic
variations, which is the main goal of this step of our project.

The new statistical parameters were tested using Monte Carlo
simulations, from which we verify that the even-statistical parameters can be
used to analyze statistical distributions in the same way as their non-even
counterparts. Many even-statistical parameters keep a strong relationship
with their counterparts that enables a comparison. The improvement in the
accuracy of statistical parameters depends of the distribution analyzed. For
many of these parameters the even-parameters display better accuracy (uniform - 7/9 of
the statistics improved with even; normal: 5:9 improved: Ceph: 8:9 improved: RR: 5:9 improved: EB: 2:9 improved.
The simulations were also used to estimate a coefficient to adjust the sample
size for each dispersion parameter to take into account  the dependence of
statistical parameters on the number of measurements. These are extremely
important to reduce the mis-selection of sources with few measurements.

Even-statistical parameters plus sample size corrections plus new model noise 
were used to propose non-correlated indices that can be used on single or multiwavelength
observations. The Strateva-modified function proposed in the present paper
provides a better model than previous functions and the sample size coefficients 
were designed for each statistical parameter account for its susceptibility 
to statistical variations. Indeed, the noise characteristics of bright and 
faint sources are better modelled by the Strateva-modified function. This is
extremely important since the single or multiwavelength analysis are only
possible using a noise model. The dispersion parameters provide similar
information but are susceptible to statistical variations that are slightly
different. However, combinations of statistical parameters tested do not
significantly improve the discrimination between variable stars and noise.
Finally, the non-correlated index was tested using the WFCAMCAL database.
The results were compared with those obtained with the standard deviation and
Strateva function. The mis-selection rate was reduced by about $520\%$ 
as result of our approach. Moreover, the correlated indices were recomputed
using  the even-mean and we also find a reduction in the mis-selection rate of
$18\%$. From all of the above, we summarize our recommendation to select variable stars from noise.

The first step of this project, where the tools and
selection criteria to discriminate variable stars from noise were studied, is
now concluded. The next step of this project will study period finding methods
and how use these methods to reduce or remove all mis-selected sources.

\section{Acknowledgements}

C. E. F. L. acknowledges a post-doctoral fellowship from the CNPq. N. J. G. C. acknowledges support from the UK Science and Technology Facilities Council. We thank Maria Ida Moretti who pointed us to the \citet[][]{Sokolovsky-2017} paper, which was crucial to perform a strict comparison with the most recent results. We also thank the reviewer for his/her thorough review and highly appreciate the comments and suggestions, which significantly contributed to improving the quality of the publication.

\bibliographystyle{aa}
\bibliography{mylib_nonc}


\end{document}